\documentclass[twocolumn,twocolappendix]{aastex631}

\usepackage{float}
\usepackage{bm}
\usepackage{amsmath, amssymb}
\usepackage{natbib}
\usepackage[multiple]{footmisc}
\defcitealias{Hannah2024}{Paper I}
\usepackage{comment}

\begin{document}

\newcommand{\newtext}[1]{\textbf{#1}}

\title{Counting the Unseen II: Tidal Disruption Event Rates in Nearby Galaxies with REPTiDE}

\author[0000-0002-7064-3867]{Christian H. Hannah}
\affiliation{Department of Physics and Astronomy, University of Utah\\ 115 South 1400 East, Salt Lake City, Utah 84112, USA}

\author[0000-0002-4337-9458]{Nicholas C. Stone}
\affiliation{Racah Institute of Physics, The Hebrew University, 91904, Jerusalem, Israel}
\affiliation{Department of Astronomy, University of Wisconsin, Madison, WI 53706, USA}

\author[0000-0003-0248-5470]{Anil C. Seth}
\affiliation{Department of Physics and Astronomy, University of Utah\\ 115 South 1400 East, Salt Lake City, Utah 84112, USA}

\author[0000-0002-3859-8074]{Sjoert van Velzen}
\affiliation{Leiden Observatory, Leiden University, Postbus 9513, 2300 RA, Leiden, The Netherlands}

\begin{abstract}

Tidal disruption events (TDEs) are a class of transients that occur when a star is destroyed by the tides of a massive black hole (MBH). Their rates encode valuable MBH demographic information, but this can only be extracted if accurate TDE rate predictions are available for comparisons with observed rates. In this work, we present a new, observer-friendly Python package called REPTiDE, which implements a standard loss cone model for computing TDE rates given a stellar density distribution and an MBH mass. We apply this software to a representative sample of 91 nearby galaxies over a wide range of stellar masses with high-resolution nuclear density measurements from \citet{Hannah2024}. We measure per-galaxy TDE rates ranging between 10$^{-7.7}$ and 10$^{-2.9}$ per year and find that the sample-averaged rates agree well with observations.
We find a turnover in the TDE rate as a function of both galaxy stellar mass and black hole mass, with the peak rates being observed in galaxies at a galaxy mass of  $10^{9.5}$~M$_\odot$  and a black hole mass of $10^{6.5}$~M$_\odot$. 
Despite the lower TDE rates inferred for intermediate-mass black holes, we find that they have gained a higher fraction of their mass through TDEs when compared to higher mass black holes. This growth of lower mass black holes through TDEs can enable us to place interesting constraints on their spins; we find maximum spins of $a_\bullet \approx 0.9$ for black holes with masses below $\sim10^{5.5}$~M$_\odot$. 

\end{abstract}


\section{Introduction} \label{sec:introduction}

When a star passes sufficiently close to a massive black hole (MBH), it will be torn apart by extreme tidal forces \citep{Hills1975}. After such an event, conservation of angular momentum will lead to the eventual circularization of the bound stellar material \citep{Rees1988}.  Although the underlying hydrodynamics of the circularization process are complex \citep{Hayasaki2013, Guillochon2014, Shiokawa2015, Hayasaki2016, Bonnerot2020, Bonnerot2021, Andalman2022, Steinberg2024}, observations indicate that the end result is a luminous, multiwavelength flare \citep{Bade1996, Gezari2006, van-Velzen2011, Bloom2011, Gezari2012, Arcavi2014, vanVelzen2016, Holoien2016, vanVelzen2020, Gezari2021} powered by a combination of shock dissipation \citep{Piran2015} and MBH accretion \citep{Rees1988, Loeb1997, Ulmer1999, Metzger2016}. The resulting electromagnetic transient is referred to as a tidal disruption event (TDE). Individually, TDEs are excellent probes of accretion physics as they present the rare opportunity to witness the formation and dissipation of an accretion disk around an MBH \citep[e.g.][]{Gezari2021}, sampling the poorly understood regime of super-Eddington accretion at early times \citep{Ulmer1999, Lodato2011, Dai2018} while exploring old questions in accretion disk stability at late times \citep{Shen2014, vanVelzen2019, Kaur2023}. The ability to ``light up" previously invisible MBHs makes these events perfect for detecting new MBHs and potentially weighing them through modeling of their light curves \citep[e.g.][]{Mockler2019, Ryu2020, Wen2020, Wen2022, Mummery2023}.

Despite allowing for the direct detection of distant and dormant MBHs, observed rates suggest just one TDE every $10^{4-5}$ years per galaxy on average \citep[e.g.][]{Donley02,vanVelzen2018,Sazonov2021,Yao2023}. This scarcity prevents us from using individual TDE detections 
as probes of MBH masses in any particular galaxy of interest. However, the rates of TDEs across different galaxy and MBH masses are likely correlated with both the MBH occupation fraction and mass distribution \citep{StoneMetzger2016}; understanding these correlations is critical for enabling future TDE science. For example, one might hope to use near-future samples of $\sim 10^{3-4}$ TDEs \citep{Bricman2020, Shvartzvald2024} to measure the abundance of intermediate mass black holes (IMBHs, with masses $M_\bullet \lesssim 10^6 M_\odot$), which can distinguish between debated high-redshift MBH formation channels \citep{Volonteri2010, Inayoshi2020}.  A practical approach to this goal requires some understanding of how TDE rates depend on MBH mass and other parameters to distinguish between otherwise degenerate scenarios (e.g. an IMBH occupation fraction of unity and a low per-galaxy IMBH TDE rate versus a low IMBH occupation fraction and a high per-galaxy IMBH TDE rate).  Therefore, the future of TDEs as MBH demographic probes necessitates the study of TDE rates and the comparison of theoretical predictions to observations.

Theoretical TDE rates are most commonly estimated by solving the orbit-averaged Fokker-Planck equation, which models the diffusion of stars through angular momentum space via collisional 2-body relaxation \citep[see e.g.][for reviews]{Merritt2013b, Stone2020}. 
In this approach, the TDE rate is defined by how often stars are scattered into the ``loss cone", which is the region of phase space (position and velocity) describing orbits that will undergo disruption.   As TDE rates can vary dramatically with changes to the nuclear stellar density profile, realistic attempts to estimate TDE rates are usually semi-empirical, calibrating stellar profiles from high-resolution {\it Hubble Space Telescope} observations of nearby galactic nuclei \citep{Magorrian1999, Syer1999, Wang2004,Vasiliev2013,StoneMetzger2016, Pfister2020}, although other, more theoretical, efforts exist too \citep{Polkas2024}.  

As outlined in \citet{Hannah2024}, hereafter \citetalias{Hannah2024}, the overall goal of our multi-paper project is to constrain MBH demographics in low-mass galaxies by comparing TDE rate predictions for model galaxy samples to observations. While those comparisons will be presented in our upcoming Paper III, we showcase here a range of astrophysically interesting intermediate results by dynamically modeling the loss cones of the sample galaxies presented in \citetalias{Hannah2024}. Given the nuclear resolution requirements of this sample, these rates represent the most accurate set of TDE rates for a sample of nearby galaxies with resolved densities on scales below $r=5$~pc.  This galaxy sample also covers a wide range of galaxy stellar mass and includes both early- and late-type galaxies making it more representative of the overall galaxy population than previous empirical TDE rate predictions.

In the process of computing TDE rates for our galaxy sample, we have developed an observer-friendly Python implementation of the usual steady-state loss cone formalism called REPTiDE, or Rate Estimation Program for Tidal Disruption Events, which we present here.  REPTiDE was designed as an easy-to-use Python module capable of computing the loss cone TDE rate for a given stellar mass distribution and central MBH. A practical aspect of this code is the ability to input discrete density profiles produced directly from surface brightness data rather than assuming some parametric model. Keeping with its observer-friendly theme, REPTiDE also contains a built-in function to measure the 3-D stellar density profile from a supplied surface brightness profile and mass-to-light ratio ($M/L$), following a procedure identical to \citetalias{Hannah2024}.

The remainder of this paper is as follows. 
The theoretical loss cone methodology we use is described in Section~\ref{sec:LC}. Details on the functionality, accuracy, and use of the REPTiDE software are described in Section~\ref{sec:reptide}.  In Section~\ref{sec:results}, we present the actual TDE rates for our galaxy sample and compare them with the most recent observed TDE rates, as well as other theoretical results. Lastly, our conclusions and plans for Paper III are discussed in Section~\ref{sec:con-future}.

\section{Loss Cone Theory} \label{sec:LC}

Loss cone theory was first developed half a century ago to study TDE rates \citep{Frank1976, Lightman1977, Cohn1978}, and in more recent times has seen a number of applications, not just to TDEs \citep{Magorrian1999, Wang2004,StoneMetzger2016} but also to other kinds of loss cone phenomena such as extreme mass ratio inspirals \citep{Hopman2005, Broggi2022, Qunbar2023}, hypervelocity stars \citep{Perets2007}, and MBH binary coalescence \citep{Milosavljevic2003}.  Subsequent refinements to the theory have also incorporated the role of asphericity \citep{Magorrian1999, Merritt2004, Vasiliev2013, Vasiliev2015, Kaur2024}, sizeable velocity anisotropies \citep{Lezhnin2015, Stone2018}, strong scatterings \citep{Weissbein2017, Teboul2024, Teboul2024b}, and repeated partial disruptions \citep{Broggi2024}, although for now we neglect these effects in our study, as their importance is challenging to quantify through photometric data alone. Below, we review the computation of theoretical TDE rates using the simplified, stationary state formalism appropriate for relaxed, quasi-spherical galactic nuclei \citet{Wang2004, StoneMetzger2016}, conditions we expect to be reasonable approximations in the lower mass galaxies of greatest interest for our study.

For a star to be tidally disrupted, it must reach a critical radius from the MBH during its orbit. This distance is referred to as the tidal radius ($r_{\rm t}$), and it depends\footnote{The combined effects of relativistic gravity and the internal structure of the star will often change the tidal radius for a full disruption by $\sim 10\%$, or at most a factor $\approx 2$ \citep{Guillochon2013, Ryu2020b}.} on the mass ($M_\star$) and radius ($R_\star$) of the star as well as the mass of the black hole ($M_\bullet$):
\begin{equation}
    r_{\rm t} = R_\star \left(\frac{M_\bullet}{M_\star}\right)^{1/3}. \label{eq:rt}
\end{equation}
For a TDE to occur, the pericenter of the star's orbit ($r_{\rm p}$) must be $\leq r_{\rm t}$.
The term ``loss cone'' comes from the geometric interpretation of this criterion, which corresponds to a cone in the velocity phase space of stellar orbits whose members will be tidally disrupted\footnote{Historically, the TDE loss cone also acquired its name in analogy to the plasma kinetic theory loss cones that plagued attempts to use magnetic mirrors as controlled fusion devices \citep{Rosenbluth1965}.}. When considering stars on highly eccentric orbits (i.e. the most likely stars to experience tidal disruption) in a spherical gravitational potential, this condition translates to a simple cut in action space.  Stars having a specific angular momentum, $J$, less than the loss-cone angular momentum, $J_{\rm LC} = \sqrt{2 {\rm G} M_\bullet r_{\rm t}}$, will suffer tidal disruption.  Near the MBH, the portion of action space with $J < J_{\rm LC}$ is devoid of stars, and TDE rates are set by the rate at which two-body scatterings can diffuse stars into the loss cone.

While this condition defines which stars will be tidally disrupted, it is agnostic as to the observability of such events. For larger black holes (or more compact stars), the disruption itself can occur within the event horizon, leaving no observable signature. The maximum MBH mass at which a star 
will be disrupted outside of the event horizon is referred to as the Hills mass \citep{Hills1975}: 
\begin{align} \label{eq:hills}
    M_{\rm H} &= \left(\frac{c^2 R_\star}{2GM_\star^{1/3}} \right)^{3/2} \\ &\approx 1 \times 10^8 M_\odot \left( \frac{R_\star}{R_\odot} \right)^{3/2} \left(\frac{M_\star}{M_\odot}\right)^{-1/2} \notag
\end{align}
Here we have set $r_{\rm t} = 2 r_{\rm g}$, the event horizon size for a non-spinning black hole (here $r_{\rm g}= GM_\bullet / c^2$ is the gravitational radius).
This Newtonian definition of the Hills mass is clearly crude (for example, it should be computed by using the minimum parabolic pericenter) but agrees surprisingly well with general relativistic calculations in the Schwarzschild metric; on the other hand, rapid MBH spin can increase the true Hills mass over the Newtonian value by almost an order of magnitude \citep{Kesden2012}.

Giant stars can be disrupted by these more massive MBHs, but their larger tidal radii and longer timescales lead to lower observability in time-domain surveys \citep{MacLeod2012, MacLeod2013}.
While MBH spin can increase $M_{\rm H}$ to allow for the disruption of main sequence stars by higher mass MBHs, these objects are, in any case, greatly outnumbered by their low-mass counterparts, which minimizes their contributions to the total TDE rate. Therefore, we do not consider any spin corrections or giant disruptions in this work. 

Under the assumption of spherical symmetry for the stellar density profile $\rho(r)$, the total potential\footnote{Throughout this paper we use the stellar dynamics convention of positive-definite potential and negative-definite kinetic energy.} of the system is described as:
\begin{equation} \label{eq:psi}
    \psi(r) = \frac{GM_\bullet}{r} + \frac{GM_{\rm enc}(r)}{r} + 4 \pi G  \int_{r}^{\infty} \rho(r')r' \,{\rm d}r',
\end{equation}
where $M_{\rm enc}(r)$ is the stellar mass enclosed within a radius of $r$. Assuming an isotropic velocity distribution for the stars, the stellar distribution function (DF) is given by an Eddington integral:
\begin{equation} \label{eq:df}
    f(\epsilon) = \frac{1}{8^{1/2}\pi^2 \left<M_\star\right>}\frac{\rm d}{{\rm d}\epsilon}\int_{0}^{\epsilon} \frac{{\rm d}\rho}{{\rm d}\psi}\frac{{\rm d}\psi}{\sqrt{\epsilon-\psi}} \, {\rm d}\psi,
\end{equation}
where $\epsilon$ is the negative of the usual specific orbital energy and $\left<M_\star\right>$ is the average stellar mass of the system. 

The time evolution of stellar distribution functions in dense environments can be understood via the collisional Boltzmann equation, which can be well approximated\footnote{The approximation lies in the neglect of strong, or large-angle, scatterings \citep{BarOr2013}.} by a Fokker-Planck equation in which uncorrelated two-body scatterings are accounted for as velocity-space diffusion coefficients.  In the spherical Kepler potential of an MBH, the local (coordinate-space) Fokker-Planck equation can be transformed into an orbit-averaged action-space Fokker-Planck equation \citep{Lightman1977}. In this picture, stellar orbits diffuse through the 2D space of actions $\{\epsilon, J\}$ until some kind of quasi-stationary state is achieved.  In the presence of a loss cone, this quasi-stationary state involves a logarithmic profile in $J$ \citep{Cohn1978}.

While the loss cone problem is thus best studied with a 2D, time-dependent diffusion equation \citep{Broggi2022}, the lack of detailed phase space information on distant galactic nuclei makes it challenging to apply such an approach to observations. Here, we will follow \citet{Magorrian1999, Wang2004, StoneMetzger2016} in applying a simplified loss cone formalism to observations of galactic nuclei.  This formalism exploits the timescale hierarchy that exists for the eccentric orbits relevant to the loss cone problem.  For such orbits, angular momentum diffusion is orders of magnitude faster than energy diffusion, so the latter process is neglected.  Furthermore, for highly eccentric orbits, loss cone refilling is dominated by a single orbit-averaged angular momentum diffusion coefficient:
\begin{equation} \label{eq:mu}
    \bar\mu(\epsilon) = \frac{2}{P(\epsilon)}\int_{0}^{r_{\rm a}} \frac{{\rm d}r}{v_{\rm r}(r)} \lim_{\mathcal{R} \to 0} \frac{\langle(\Delta \mathcal{R})^2\rangle}{2\mathcal{R}}.
\end{equation}
Here, $\mathcal{R} \equiv J^2/J_{\rm c}^2$ where $J_{\rm c}(\epsilon)$ is the angular momentum of a circular orbit and $r_{\rm a}$ specifies the apocenter radius for a star with energy $\epsilon$. The orbital period, $P(\epsilon)$, is defined as $\int_{0}^{r_{\rm a}(\epsilon)}{{\rm d}r/\sqrt{2(\psi - \epsilon)}}$.

In equation \ref{eq:mu}, the orbit-averaged diffusion coefficient $\bar{\mu}$ is computed from the $\mathcal{R}\to0$ limit of a local angular momentum diffusion coefficient, which can be expressed as:
\begin{align} \label{eq:lim}
    \lim_{\mathcal{R} \to 0} \frac{\langle(\Delta \mathcal{R})^2\rangle}{2\mathcal{R}} = &\frac{32 \pi^2 r^2 G^2 \langle M_\star^2 \rangle {\rm ln}\Lambda}{3 J_{\rm c}^2(\epsilon)} \\
    &\times \left(3 I_{1/2}(\epsilon)-I_{3/2}(\epsilon)+2 I_{0}(\epsilon) \right). \notag
\end{align}
Here, the $I_{n}$ terms encode combinations of local velocity-space diffusion coefficients (see e.g. \citealt{Merritt2013} for more details) and can be practically computed as moments of the DF:
\begin{equation}
    I_0(\epsilon) \equiv \int_{0}^{\epsilon}f(\epsilon'){\rm d}\epsilon
\end{equation}
\begin{align}
    I_{n/2}(\epsilon) \equiv ~&[2(\psi(r)-\epsilon)]^{-n/2}  \\
    &\times \int_{\epsilon}^{\psi(r)}[2(\psi(r)-\epsilon')]^{n/2} f(\epsilon'){\rm d}\epsilon. \notag
\end{align}
Additionally, ${\rm ln}\Lambda \approx {\rm ln}(0.4 M_\bullet/M_\star)$ is the Coulomb logarithm and $\langle M_\star^2 \rangle$ is the second moment of the stellar present day mass function (PDMF):
\begin{equation} \label{eq:m2}
    \langle M_\star^2 \rangle \equiv \int \frac{{\rm d}N_\star}{{\rm d}M_\star}M_\star^2 {\rm d}M_\star.
\end{equation}
The diffusion coefficient $\bar{\mu}$ is ultimately a weighted average over scatterings from a population of stars, and the contribution of different stellar mass bins is captured by $\langle M_\star^2 \rangle$. 
With these definitions, the flux of stellar scattering into the loss cone  (i.e. the number of stars experiencing tidal disruption) per unit time and energy can be expressed as:
\begin{equation} \label{eq:lcflux}
    \mathcal{F}(\epsilon){\rm d}\epsilon = 4 \pi^2 J_{\rm c}(\epsilon) \bar\mu(\epsilon) \frac{f(\epsilon)}{{\rm ln}\mathcal{R}_0^{-1}} {\rm d}\epsilon, 
\end{equation}
where $R_0(\epsilon)$ defines the minimum populated angular momentum, below which no stars exist due to disruption. This can be approximated \citep{Cohn1978} as:
\begin{align}
    \mathcal{R}_0(\epsilon) = ~&\mathcal{R}_{\rm LC}(\epsilon) \\
    &\times
    \begin{cases}
    \exp (-q)&, q>1 \\ 
    \exp (-0.186q-0.824q^{1/2})&, q<1.
    \end{cases} \notag
\end{align}
Here, $q$ is a dimensionless diffusivity, the ratio of the per-orbit change in $\mathcal{R}$ to the loss cone value $\mathcal{R}_{\rm LC}$: 
\begin{equation}
    q(\epsilon) = \bar{\mu}(\epsilon) \frac{P(\epsilon)}{\mathcal{R}_{\rm LC}(\epsilon)}.
\end{equation}
When $q \ll 1$, we are in the ``empty loss cone'' regime: $\mathcal{R}_0 \approx \mathcal{R}_{\rm LC}$, and stars slowly diffuse through $J$-space until they are destroyed in a grazing disruption (generally; see \citealt{Weissbein2017}).  In this regime, the TDE rate is determined by $\bar{\mu}$.  Conversely, when $q \gg 1$, we are in the ``full loss cone" regime (also referred to as the pinhole regime): $\mathcal{R}_0 \ll \mathcal{R}_{\rm LC}$, and stars may wander in and out of the loss cone many times in a particular orbit.  In this regime, the TDE rate is independent of $\bar{\mu}$ and is determined entirely by the DF.  In realistic galaxies, small radii (high $\epsilon$) have empty loss cones and large radii (low $\epsilon$) have full loss cones.

REPTiDE provides a simple, user-friendly Python implementation of this theory, which we will now discuss further in Section~\ref{sec:reptide}.


\section{A new loss-cone software package: REPTiDE} \label{sec:reptide}
We present REPTiDE (or Rate Estimation Program for Tidal Disruption Events), a new publicly available software package\footnote{\url{https://github.com/christianhannah/reptide}} that implements the loss cone physics from Section~\ref{sec:LC} as a series of functions written in Python and computes the expected TDE rate for a spherically symmetric galactic nucleus. 
In this section, we describe the functionality of REPTiDE along with its key features. 

The primary inputs required for this calculation are the central black hole mass ($M_\bullet$) and the radial 3D stellar density profile $\rho(r)$. There are two versions of the code that depend on the definition of the stellar density profile; this can be either ``Discrete" or ``Analytic."  In the discrete version, the density profile is supplied as a discrete, non-parametric set of data points. The analytic version uses a density profile that follows a power-law with exponential decay:
\begin{align}
    \rho(r) = ~ &\rho_{5{\rm pc}} \left(\frac{r}{5~{\rm pc}}\right)^{-\gamma} \\
    &\times
    \begin{cases}
    1&, r < r_{\rm d} \\ 
    \exp [-(r-r_{\rm d})/\sigma_{\rm d}]&, r \geq r_{\rm d}.
    \end{cases} \notag
\end{align}
Here, $\rho_{5{\rm pc}}$ is the stellar density at a radius of 5 pc, $\gamma$ is the power-law slope, $r_{\rm d}$ is the radius where the exponential decay starts, and $\sigma_{\rm d}$ is the width of the decay. The analytic version is designed to take inputs from the galaxy stellar mass vs. density and power-law slope relations presented in \citetalias{Hannah2024}.  This version of input was developed for our planned application of REPTiDE to model galaxy samples in Paper III (Hannah et al., {\em in prep}).

\subsection{1D Surface-Brightness Profile to 3D Density}
3D stellar densities are not a directly observed quantity, and calculating them from an observed brightness profile requires some effort and assumptions. For this reason, we have included a function in REPTiDE dedicated to this conversion. Following an identical procedure to \citetalias{Hannah2024}, this function (``\_reptide\_.SB\_to\_Density()") uses a multi-Gaussian expansion (MGE) fit to the supplied 1D surface-brightness profile via the ``mgefit" package of \citet{Cappellari2002-MGE}. This MGE fit is then converted to a 3D stellar density profile using the user-supplied mass-to-light ($M/L$) ratio. The returned density profile is spherically averaged, and thus, the user may specify the inclination angle and observed axial ratio ($b/a$ where $a$ and $b$ are the lengths of the semimajor and semiminor axes, respectively) of the galactic nucleus. If this information is unknown, an axial ratio of 1 and inclination of $90^{\circ}$ are assumed.

\subsection{Procedure} \label{sec:procedure}

The calculation begins by computing the gravitational potential defined in Eq.~\ref{eq:psi}, which is subsequently used to define the range of specific orbital energies to consider.   The boundaries on orbital energies are defined as the potential at 100~$r_{\rm t}$ (for a Solar mass star) and the potential at $10^6$ pc. These boundaries define a wide range in energy space and ensure the peak of the loss-cone flux curve is captured. The default energy grid consists of 1000 energies, but this can be altered with the ``n\_energies" keyword, which specifies the desired number of logarithmically spaced orbital energies. However, we do not encourage using a small value for ``n\_energies," for reasons quantified later in Section~\ref{sec:orben_res}.

The code then computes the stellar DF (Eq. \ref{eq:df}) using a double exponential transform to handle the integrable singularity in the DF integral. The next step is to compute the following quantities as functions of orbital energy: orbit-averaged angular momentum diffusion coefficient ($\bar\mu$), period ($P$), the squared ratio of the loss-cone angular momentum to the angular momentum of a circular orbit ($\mathcal{R}_{\rm LC}$), and the diffusivity parameter ($q$). As with the DF integral, we use a double exponential transform for the edge singularity in the period integral. These quantities are first calculated for a ``monochromatic'' population of 1~M$_\odot$ stars and are later adjusted when accounting for the present-day stellar mass function (see below). With all of the above functions defined, the loss-cone flux (Eq.~\ref{eq:lcflux}) is tabulated and integrated over all orbital energies to define the ``Solar" TDE rate. 

Monochromatic stellar populations do not exist in reality, and thus, the final step in this calculation involves accounting for the present-day stellar mass function (PDMF). Implementing a realistic PDMF has two competing effects on the TDE rate. The inclusion of more sub-Solar mass stars increases the TDE rate due to their large numbers, but this can also change the angular momentum diffusion coefficient (Eq. \ref{eq:mu}), altering the TDE rate. \citet{StoneMetzger2016} tested two different PDMFs (Salpeter and Kroupa), and both were shown to have similar rate enhancements ($\approx 1.63$ for Salpeter and $\approx 1.53$ for Kroupa with $M_\star^{\rm min} = 0.08$ M$_\odot$ and $M_\star^{\rm max} = 1$ M$_\odot$). In their investigation, $M_\star^{\rm max}$ had a larger effect on the level of rate enhancement (which is directly correlated with the age of the stellar system) because the diffusion coefficient is set by the most massive stars present. REPTiDE implements the Kroupa PDMF \citep{Kroupa2001} below:
\begin{align}
&\left.\frac{dN_{\star}}{dM_{\star}}\right. = \\
&\begin{cases}
0.98(M_\star/M_{\odot})^{-1.3}&, M_\star^{\rm min}<M_\star<0.5 M_\odot \\
2.4(M_\star/M_{\odot})^{-2.3}&, 0.5M_\odot<M_\star<M_\star^{\rm max} \\
0&, {\rm otherwise},
\label{eq:Kroupa}
\end{cases} \notag
\end{align}
where $M_\star^{\rm min}$ and $M_\star^{\rm max}$ specify the minimum and maximum stellar masses to consider. These values can be set using the {\it M\_min} and {\it M\_max} parameters, respectively. 

To determine the ``full" TDE rate, the PDMF is first defined over the specified mass range and normalized. This PDMF is then used to compute $\left< M_\star \right>$ and $\left< M_\star^2 \right>$, which are set to M$_\odot$ and M$_\odot^2$ in the monochromatic case, respectively. Each stellar mass considered possesses a different tidal radius ($r_{\rm t}$), and thus, some quantities have to be adjusted to compute the TDE rate per stellar mass. First, $\bar\mu(\epsilon)$ must be adjusted given the new value for $\left< M_\star^2 \right>$ and the new integral bounds (as we set $r_{\rm p}$ to $r_{\rm t}$ in the orbit-averaging calculations). We also adjust $\mathcal{R}_{\rm LC}(\epsilon)$ for the new value of $r_{\rm t}$. These quantities then define the adjusted $q(\epsilon)$ values needed to compute the loss-cone flux curve $\mathcal{F}(\epsilon){\rm d}\epsilon$ for a given stellar mass. This process is performed for each of the 50 stellar masses, and the integrated loss-cone flux curves give the TDE rate per stellar mass. Lastly, this TDE rate per stellar mass is multiplied by the PDMF and integrated to give the ``full" TDE rate. 

When computing the ``full" TDE rate, REPTiDE can account for TDE rate suppression due to direct capture, which is enabled through the ``{\it EHS}" flag. When set, the final integration of the TDE rate per stellar mass is only computed for stellar masses with a Hills mass less than the MBH (as defined by Equation~\ref{eq:hills}). All TDE rates presented in this work are ``full" rates with minimum and maximum stellar masses of $0.08$~M$_\odot$ and $1$~M$_\odot$, respectively.

\subsection{Handling Small Radii} \label{sec:small_radii}
In these calculations, extrapolation of the density profile to small radii (well inside the resolution of the density measurements) is sometimes required. Extrapolating very steep power-law density profiles can cause a few issues: if $\gamma \ge 2$, diffusion of stars through energy space becomes strongly non-local; if $\gamma \ge 9/4$, the TDE rate actually diverges as one considers more tightly bound stars; if $\gamma \ge 3$, the stellar mass enclosed diverges. However, it is generally unphysical for nuclear density profiles to be characterized by power-laws with very large $\gamma$ values over an arbitrary range of radii \citep{Stone2018}. To overcome these issues, we have implemented the ability to set a Bachall-Wolf cusp density profile for the innermost radii. 

The Bachall-Wolf cusp is a theoretically expected steady-state distribution of stars around an MBH formed through collisional relaxation processes (\citealt{Bahcall1976, Bahcall1977}). When $\gamma>3/2$, relaxation times are the shortest near the MBH, so this cusp is expected to form from the inside out, resulting in a broken power-law density profile with a break radius related to the dynamical age of the system \citep{Stone2018}. Main sequence stars in relaxed stellar cusps usually follow a power-law with $\gamma_{\rm BW}\approx 1.5-1.75$, where the exact value depends on the distribution of stellar masses \citep{Alexander2009, Preto2010}. 

\begin{figure}[!t]
    \centering
    \includegraphics[width=0.9\linewidth]{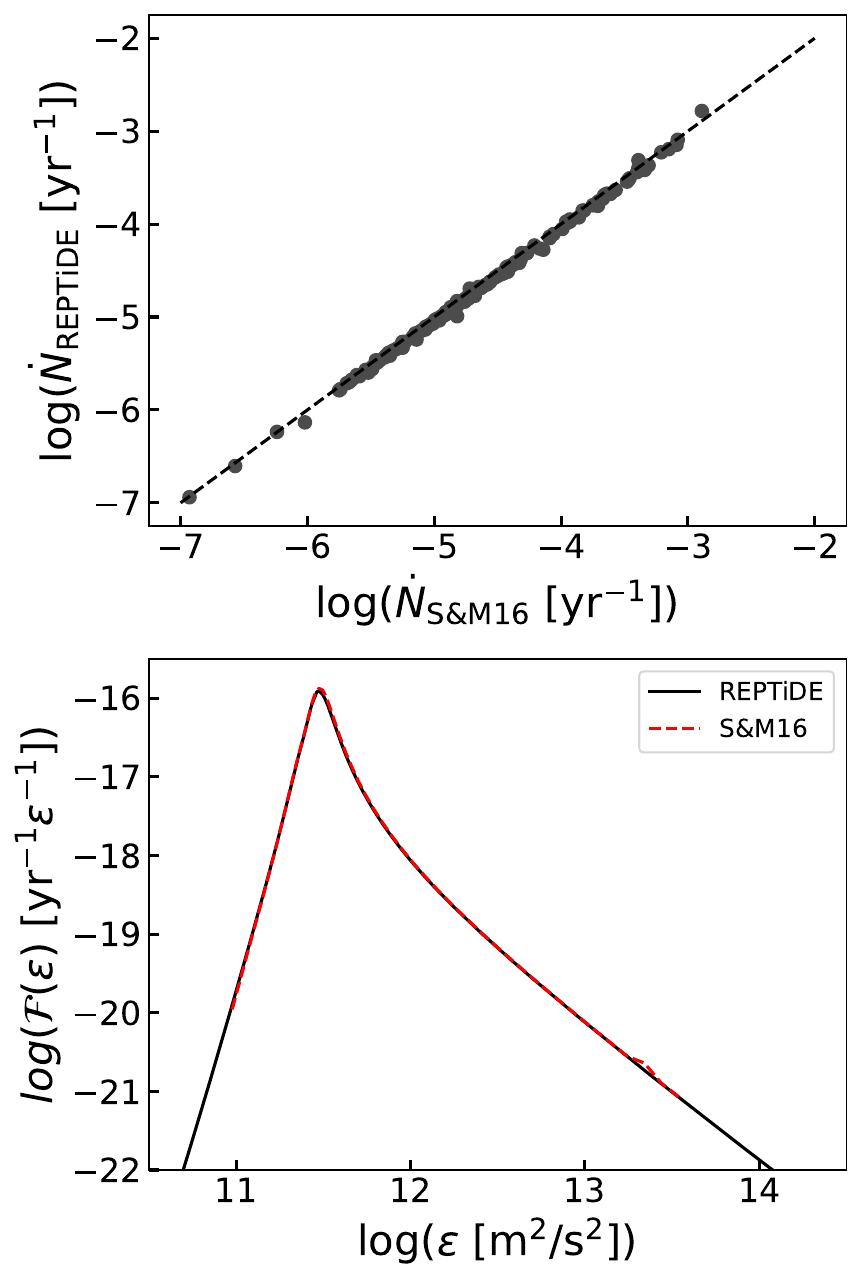}
    \caption{Top: One-to-one comparison of TDE rates for the \citet{StoneMetzger2016} galaxy sample with REPTiDE using identical density profiles and black hole masses (dashed line shows the one-to-one line). Bottom: Loss-cone flux curve for NGC~4551 from \citet{StoneMetzger2016} (red dashed) compared to the same flux curve computed by REPTiDE (black).}
    \label{fig:accuracy}
\end{figure}

An approximate relation between the relaxation radius (i.e. the radial extent of the Bachall-Wolf cusp) and the dynamical age of the system can be derived by setting the relaxation time equal to the dynamical age. For a power-law density profile described as $\rho(r) = \rho_{\rm 5pc} \left(\frac{r}{{\rm 5~pc}}\right)^{-\gamma}$, this relation is:
\begin{equation} \label{eq:t_age}
    t_{\rm relax} \approx \frac{\kappa [G(M_{\bullet}+M(r_{\rm relax}))]^{3/2} \left< M_{*}\right> r_{\rm relax}^\gamma}{G^2 \rho_{\rm 5pc} (5~{\rm pc})^\gamma \left< M_{*}^2\right> r_{\rm relax}^{3/2} (1+\gamma)^{3/2} {\rm ln}\Lambda},
\end{equation}
where $\kappa = 0.34$ \citep{Binney2008} and $M(r_{\rm relax}) = 4\pi (3-\gamma)^{-1} \rho_{\rm 5pc} (5~{\rm pc})^\gamma r_{\rm relax}^{3-\gamma}$ is the stellar mass contained within the relaxation radius.

When applying REPTiDE to the sample galaxies from \citetalias{Hannah2024} (see Section~\ref{sec:results}), extrapolation of the density profiles inward of our resolution limits is required. Thus, for all galaxies with density profiles steeper than a Bachall-Wolf cusp ($\gamma\geq1.75$), we use Equation~\ref{eq:t_age} to estimate the relaxation radius under the conservative assumption that the system has relaxed over a Hubble time. For a few galaxies, this assumption is inconsistent as a Bachall-Wolf cusp is expected to be resolved. In these cases, we apply the cusp at the resolution limit. We note that the TDE rate is quite sensitive to the value of $r_{\rm relax}$ with the effect becoming largest for the steepest density profiles. This uncertainty is discussed in more detail in Section~\ref{sec:errors}.

\subsection{Accuracy}

To ensure the accuracy of REPTiDE's calculations, we perform direct comparisons with the output from \citet{StoneMetzger2016} (Figure~\ref{fig:accuracy}). The top panel shows a one-to-one comparison of the TDE rates computed by REPTiDE using identical density profiles and black hole masses to those used in \citet{StoneMetzger2016}\footnote{These are not the final TDE rates presented for our sample galaxies.}. The median rate residual from this investigation is 0.019 dex and is believed to be caused by minor differences in orbital energy grid densities (described further in Section~\ref{sec:orben_res}). Overall, the rates agree very well. To further highlight the agreement, the lower panel shows the loss-cone flux curve for a galaxy from \citet{StoneMetzger2016} where the density extrapolation was handled identically by 
the non-public code used for \citet{Stone2016} and REPTiDE. These results together highlight the accuracy of the computations performed by this software.

\subsection{Effect of Orbital Energy Resolution} \label{sec:orben_res}
The number of orbital energies considered for each calculation is directly related to the execution speed of REPTiDE. Thus, we performed a convergence test to gauge the best resolution for the orbital energies while maximizing efficiency. This test involved calculating the TDE rate for NGC 4551 (same galaxy in Figure~\ref{fig:accuracy}) with 10 logarithmically spaced numbers of orbital energies from 100 to 100,000. We find that if the energy grid is too coarse, the resulting TDE rate is always higher than the finer grids. For instance, the TDE rate computed with 50 energies was $0.79$~dex larger than the finest grid. We select a default value of orbital energies of 1000 in this work, as this results in the TDE rate being less than 1\% off from even higher values of energy resolution. We provide further details on our convergence tests regarding energy binning in Appendix \ref{app:converge}.

\section{TDE Rates for Real Galaxy Sample} \label{sec:results}

\begin{figure*}[!t]
    \centering
    \includegraphics[width=0.9\linewidth]{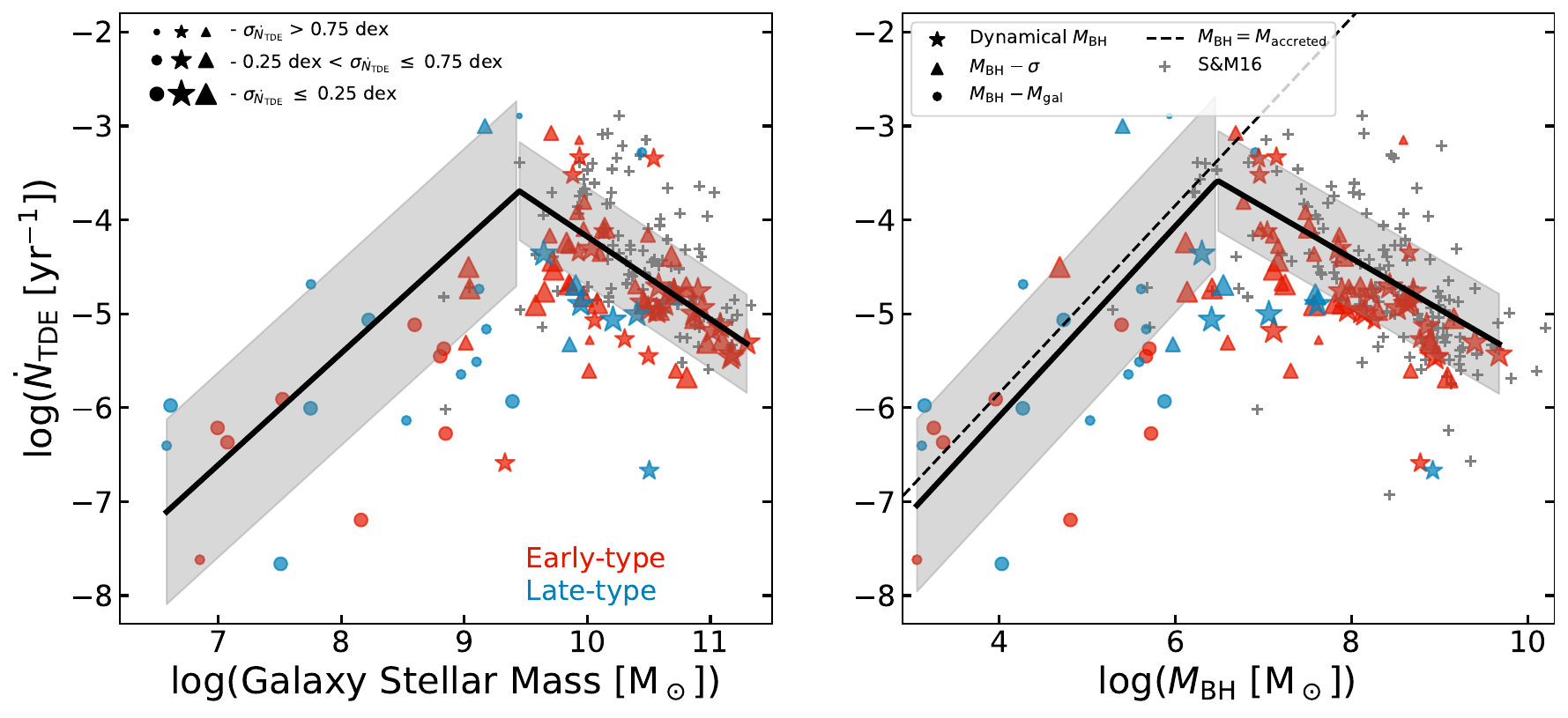}
    \caption{Per-galaxy TDE rates for our sample galaxies as functions of galaxy mass (left) and MBH mass (right), colored by galaxy-type. The symbol sizes specify the uncertainty on the rates and the shape indicates the origin of the MBH mass measurement. The solid black lines give the best-fit broken power-laws to the data with the scatter (derived separately above and below the break radius) shown as the gray shaded regions. The published results from \citet{StoneMetzger2016} are shown as gray crosses in both panels. Note that the results in this figure do not include the event horizon suppression that limits the number of observable TDEs at high black hole masses.  The black dashed line in the right panel gives the maximum MBH mass allowed assuming a constant TDE rate over a Hubble time.  If galaxies could sustain rates above this line for a Hubble time, their MBH masses would grow through stellar consumption until they moved to the right of the line. The lack of galaxies above or even near the line suggests a subdominant role for TDEs in MBH growth, even at the low-mass end, albeit with a few outliers.}
    \label{fig:rates_v_mbh}
\end{figure*}

In \citetalias{Hannah2024}, we constructed a sample of 91 galaxies where 3D stellar densities could be reliably measured on pc-scales. This sample is focused primarily on nucleated galaxies (i.e. ones hosting NSCs at their centers) extending down to stellar masses of 10$^{6.58}$~M$_\odot$ and was used to construct scaling relations between NSC density structure and galaxy mass.  While this sample is not a volume-limited sample, it includes a range of Hubble types and includes galaxies selected from distance-limited samples \citep{Pechetti2020,Hoyer2023a}, making these galaxies representative of those in the local Universe.  In this work, we present the theoretically expected TDE rate for each galaxy in this sample computed with REPTiDE.

For these calculations, we use the best available central MBH mass measurement for each galaxy along with the discrete density profiles (i.e. using the discrete version of REPTiDE). The MBH masses were derived in one of three methods (in decreasing order of reliability): (1) Dynamical mass measurements \citep[available for 30 galaxies;][]{Greene2020,van-den-Bosch2016, Reines2015}, (2) A galaxy morphology dependent M$_{\rm BH}$-$\sigma$ relation from \citet{Greene2020}, and (3) A galaxy morphology dependent M$_{\rm BH}$-M$_{\rm gal}$ relation from \citet{Greene2020} (see \citetalias{Hannah2024} for more details). We fix the slope used in the inward extrapolation of the density profiles to the 3D power-law slopes derived in \citetalias{Hannah2024}, unless the nucleus is steeper than a Bachall-Wolf cusp ($\gamma < -1.75$). As discussed in Section~\ref{sec:small_radii}, these systems are not stable and should relax into such a cusp. This expectation, paired with diverging TDE rates resulting from ``ultrasteep" power-law density slopes ($\gamma < -2.25$), motivates our decision to apply a Bachall-Wolf cusp slope of $-7/4$ inward of our resolution limit for these galaxies (47/91 galaxies). The radial extent of the cusps in these galaxies is defined through Equation~\ref{eq:t_age}, assuming relaxation over a Hubble time (unless this suggests a relaxation radius larger than the resolution limit, in which case we apply the cusp at the resolution limit; this occurs in 12/47 galaxies). For large radii beyond our measurements, we apply a simple exponential decay with a width of $1$~kpc. 

Figure~\ref{fig:rates_v_mbh} shows the TDE rates for these galaxies as functions of galaxy stellar mass (left panel) and black hole mass (right panel). These purely dynamical rate predictions do not account for event horizon rate suppression (which we will return to later) and assume a Kroupa PDMF with masses spanning from 0.08~M$_\odot$ to 2~M$_\odot$, which represents a $\sim1$~Gyr old stellar population. While there is some uncertainty introduced by unknown nuclear stellar ages, the PDMF enhancement factor ranges from $1.5$ to $3.8$ for stellar ages of $10$~Gyr to $30$~Myr \citep{StoneMetzger2016}. Uncertainties in the TDE rates are indicated by symbol size, with the least certain being smaller and vice versa. The symbol shapes indicate the origin of the $M_\bullet$ estimate. Both of these plotting schemes are used for all results figures in this paper. 

Previous TDE rate measurements from \citet{StoneMetzger2016} are shown as gray crosses. Although the rates from \citet{StoneMetzger2016} are higher on average, there is considerable overlap in these results. For the 30 galaxies belonging to both samples, all have higher $M/L$-ratios in \citet{StoneMetzger2016}, which is the most likely reason for the higher rates, as increasing the $M/L$ will increase the overall stellar density at all radii. It is also worth noting here that none of the 150 density profiles used in \citet{StoneMetzger2016} require Bachall-Wolf cusps at small radii ($\gamma \geq 2.25$, of which we have 26), which further highlights the importance of resolving NSC scales when computing TDE rates. This is likely due to the use in \citet{StoneMetzger2016} of parametrized ``Nuker-law'' surface brightness profiles \citep{Lauer2005, Lauer2007} that were fit over a relatively large range of spatial scales and were therefore less sensitive to steep slopes produced by nuclear star clusters.

To characterize these relations and aid future comparisons, we fit broken power-laws to this data weighted by the TDE rate uncertainties, which are shown as solid black lines in Figure~\ref{fig:rates_v_mbh} and described as:

\begin{align} \label{eq:bpl_fit_eq}
    &{\rm log}(\dot N_{\rm TDE}) = \\
    &\begin{cases}
    {\rm log}(A) + \alpha*{\rm log}\left( \frac{M_{\rm BH/gal}}{M_{\rm norm}}\right),  \, M_{\rm BH/gal} < M_b\\ 
    {\rm log}(B) + \beta*{\rm log}\left( \frac{M_{\rm BH/gal}}{M_{\rm norm}}\right), \, M_{\rm BH/gal} \geq M_b.
    \end{cases} \notag
\end{align}

Here, $M_{\rm BH/gal}$ represents the MBH mass or galaxy stellar mass (depending on the desired relation), $M_b$ is the break mass, $\alpha$ and $\beta$ are the slopes, and the normalization constants ($A$ and $B$) are related by: $A = B - {M_b/{\rm M}_{\rm norm}}^{\alpha-\beta}$. Because the MBH and galaxy mass ranges differ, we normalize each power law at a different mass specified by $M_{\rm norm}$; the galaxy mass fit uses $M_{\rm norm} = 10^9$~M$_\odot$ while the MBH mass fit uses $M_{\rm norm} = 10^6$~M$_\odot$. For TDE rates as a function of galaxy mass, we find: ${\rm log}(A) = -4.23 \pm 0.42$, $\alpha = 1.19 \pm 0.27$, $\beta = -0.88 \pm 0.48$, and ${\rm log}(M_b) = 9.45 \pm 0.37$. Our best-fit parameters for $\dot N$ as a function of MBH mass are: ${\rm log}(A) = -4.07 \pm 0.45$, $\alpha = 1.01 \pm 0.23$, $\beta = -0.54 \pm 0.23$, and ${\rm log}(M_b) = 6.48 \pm 0.50$.  Note that all logarithms in this paper are base-10.

The uncertainties on fit parameters were measured via 1000 iterations of bootstrap resampling (as are all fit parameter errors presented in this work). We estimate the scatter in these fits using Gaussian deconvolution of the fit residuals assuming the residuals follow a Gaussian distribution as in \citet{Pryor1993}. We include the TDE rate uncertainties here as the heteroscedastic errors on the fit residuals. The scatter was measured both above and below the break mass and is shown as the gray-shaded regions around the best-fit broken power law; at low masses, the scatter is 0.98/0.91 dex ($M_{\rm gal}/M_\bullet$), while at higher mass, it is 0.52/0.53 dex. 

The black dashed line in Figure \ref{fig:rates_v_mbh} shows the maximum black hole mass given a constant TDE rate over a Hubble time, assuming each TDE contributes $\left<M_\star\right>/2$. If a galaxy were to lie above this line, its position would indicate that it could not have maintained its present-day mass and TDE rate for a Hubble time (as otherwise stellar consumption would have dramatically increased the mass of the MBH, moving its position below/to the right of this line).  This line can be viewed as a crude self-consistency check in our TDE rate calculation, at least at the ensemble level, however, one can always invoke an unusual evolutionary history for any individual outlier.  

Almost all galaxies in our sample do indeed fall below this line, suggesting that the uncertainties in our modeling (most notably, the extrapolation of galaxy scaling relations into the IMBH regime to estimate black hole masses) are not causing dramatic errors.  The nine galaxies above this line could arise from variable TDE rates over the life of these galaxies (i.e. with lower TDE rates at earlier times), a misestimation of stellar mass-to-light ratios, or an underestimate of $M_\bullet$.  The lack of a galaxy pile-up just below this line hints that for most galaxies in our sample, TDEs have not dominated MBH growth \citep{Stone2017}, though reaching a firm conclusion in this regard would require more robust IMBH mass estimates.

\subsection{Comparisons with Literature}

\begin{figure*}[!t]
    \centering
    \includegraphics[width=0.9\linewidth]{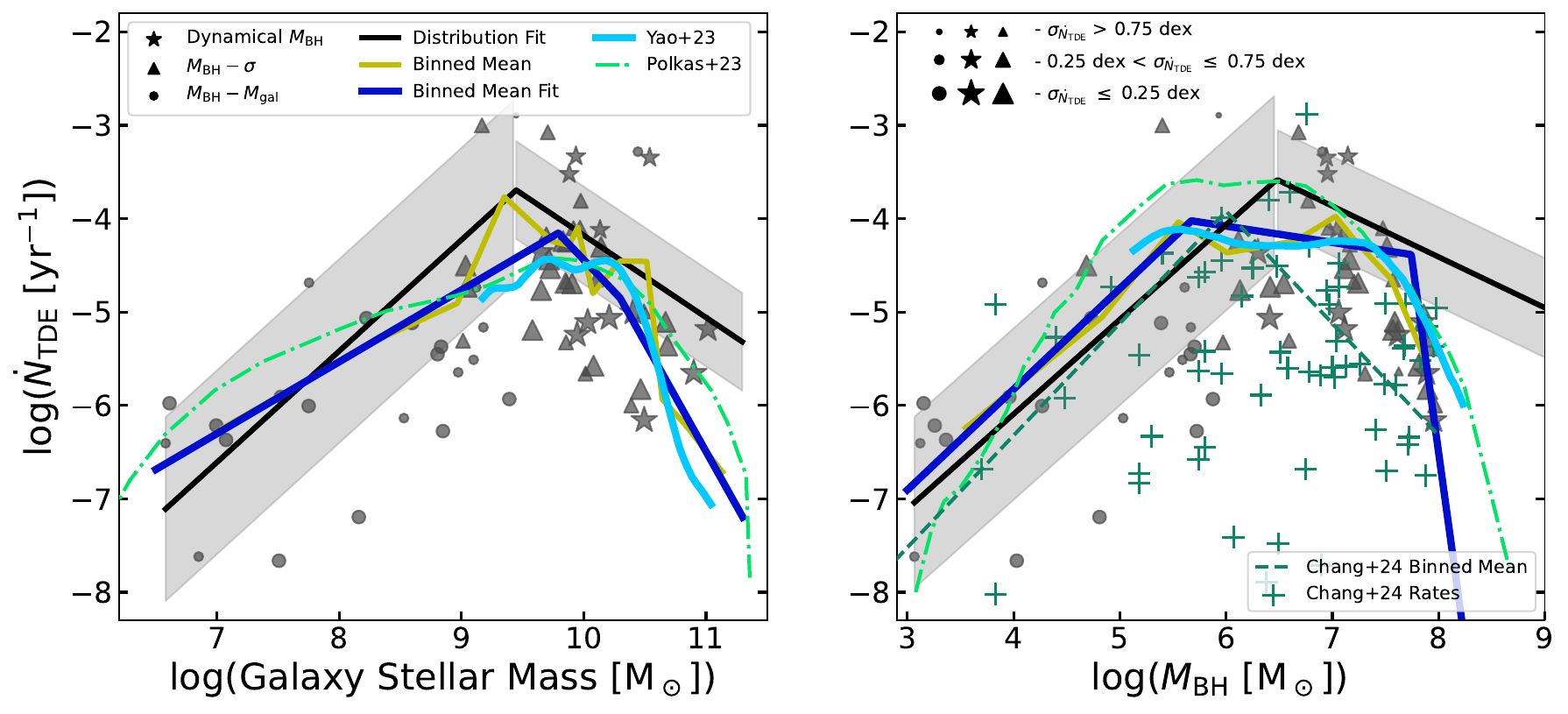}
    \caption{Similar to Figure~\ref{fig:rates_v_mbh}, but here we compare our TDE rates to the most recent observed TDE rate distributions from \citet[][cyan]{Yao2023}.
    The solid black lines are the same as Figure~\ref{fig:rates_v_mbh} and describe the typical TDE rates expected for different galaxy and MBH masses. Given that observed rates will be dominated by the highest TDE rate galaxies in each mass bin, we show a binned average TDE rate in blue for comparison. Other theoretical TDE rate results from \citet[][green dot-dashed]{Polkas2024} and \citet[][green crosses and dashed]{Chang2024} are overplotted for comparison as well.}
    \label{fig:rates_v_mbh_comp}
\end{figure*}

Tension has existed between observed and theoretical TDE rates for years. Previous loss cone-based TDE rates from \citet{Magorrian1999,Wang2004,StoneMetzger2016} have over-predicted the expected TDE rates for Milky Way-like galaxies by a factor of a few to an order of magnitude when compared to rate inferences from time domain surveys. Most past loss cone modeling predicted a rate of $\sim$ $\times 10^{-4} {\rm gal}^{-1}{\rm yr}^{-1}$, while observed rates inferred from optical and X-ray surveys \citep[e.g.][]{Donley02,vanVelzen2014,van-Velzen2020,Yao2023} usually suggest a rate of $\sim$ few $\times 10^{-5} {\rm gal}^{-1}{\rm yr}^{-1}$. 

The resolution of this tension is not clear at present. While most assumptions of standard loss cone modeling are conservative \citep{StoneMetzger2016}, there are dynamical mechanisms to reduce TDE rates. Most notably, a tangential velocity anisotropy, arising either from the aftermath of an MBH merger \citep{Lezhnin2015} or from loss cone ``shielding'' by a steep cusp of stars and/or stellar mass black holes \citep{Teboul2024, Teboul2024b}, will be capable of dramatically suppressing rates relative to the quasi-isotropic cusps we consider here.  On the other side of the equation, observationally inferred rates are sensitive to other uncertainties; the poorly characterized bottom end of the TDE luminosity function \citep{vanVelzen2018}, nuclear dust obscuration \citep{Roth2021}, and uncertainties in TDE emission mechanisms \citep{Dai2018,vanVelzen2021}.  In this work, however, we re-examine this basic tension using our larger galaxy sample, with its relatively careful treatment of central light profiles.

Figure~\ref{fig:rates_v_mbh_comp} compares our TDE rates to the most recent observed rates from \citet[][light blue lines]{Yao2023} in addition to other recent theoretical rate results. Unlike Figure~\ref{fig:rates_v_mbh}, the TDE rates presented here account for event horizon suppression as this dramatically affects the observability of TDEs. We use the same mass functions as \citet{Yao2023}, namely the \citet{Baldry2012} galaxy stellar mass function and the black hole mass function from \citet{Gallo2019}, to convert the observed volumetric rates to per-galaxy rates. 

Our fits to the distributions of TDE rates from Figure~\ref{fig:rates_v_mbh} (black line and grey shaded band) are not ideal for comparison to observed rates for two reasons. First is the lack of event horizon suppression at the high mass end. Secondly, at a given galaxy or MBH mass, galaxies will span a wide range of TDE rates and observed rates will be dominated by the highest TDE rate galaxies, even when subdominant in the overall population.
Thus, we perform additional fits to the binned means of our rate results including event horizon rate suppression via direct capture (yellow lines show the binned means and blue lines show the fits; Figure~\ref{fig:rates_v_mbh_comp}). Given that the Hills mass has a clear physical origin and is separate from a break in TDE rates near the IMBH-SMBH transition, it is important to capture both of these effects separately. As such, we fit doubly broken power laws to this data described by: 
\begin{align} \label{eq:dbl_bpl_fit_eq}
    &{\rm log}(\dot N_{TDE}) = \\
    &\begin{cases}
    {\rm log}(A) + \alpha*{\rm log}\left( \frac{M_{\rm BH/gal}}{M_{\rm norm}}\right),  \, M_{\rm BH/gal} < M_{b,1}\\ 
    {\rm log}(B) + \beta*{\rm log}\left( \frac{M_{\rm BH/gal}}{M_{\rm norm}}\right), \, M_{b,1} \leq M_{\rm BH/gal} < M_{b,2}\\
    {\rm log}(C) + \zeta*{\rm log}\left( \frac{M_{\rm BH/gal}}{M_{\rm norm}}\right), \, M_{\rm BH/gal} \geq M_{b,2}.
    \end{cases} \notag
\end{align}
Similarly to the single broken power law fits in Figure~\ref{fig:rates_v_mbh}, $M_{\rm BH/gal}$ specifies the MBH or galaxy mass, and $M_{\rm norm}$ gives the normalization mass (same as previous fits). Here, the normalization constants are related by:\\ $A = B - {M_{b,1}/{\rm M}_{\rm norm}}^{\alpha-\beta}$ and $C = B - {M_{b,2}/{\rm M}_{\rm norm}}^{\beta-\zeta}$.

To avoid introducing uncertainties from uneven numbers of galaxies in fixed bin sizes \citep[e.g.][]{Maiz-Apellaniz2005}, we implement variable bin sizes ensuring 7 galaxies per bin. Uncertainties on the TDE rates were used to perform a weighted mean in each bin with the error on the mean defined by adding the individual rate uncertainties in quadrature.
We find the following best-fit parameters for the binned mean TDE rates:\\ {\bf $\dot N~{\rm vs.}~M_{\rm gal}$}:~ ${\rm log}(A) = -4.76 \pm 0.48$, $\alpha = 0.77 \pm 0.28$, $\beta = -1.376 \pm 1.08$, $\zeta = -2.34 \pm 2.36$, ${\rm log}(M_{b,1}) = 9.79 \pm 0.25$, ${\rm log}(M_{b,2}) = 10.30 \pm 0.71$; \vspace{5pt} \\ 
{\bf $\dot N~{\rm vs.}~M_\bullet$}:~ ${\rm log}(A) = -3.67 \pm 0.73$, $\alpha = 1.08 \pm 0.31$, $\beta = -0.18 \pm 0.81$, $\zeta = -8.24 \pm 3.89$, ${\rm log}(M_{b,1}) = 5.67 \pm .54$, ${\rm log}(M_{b,2}) = 7.74 \pm .54$.  
\vspace{5pt}

Across the full range of galaxy masses, the left panel of Fig.~\ref{fig:rates_v_mbh_comp} shows our average rates agree reasonably well with observed TDE rates from ZTF \citep{Yao2023}. For the first time, our theoretical rate predictions for Milky Way mass galaxies 
are consistent with observational estimates of a few $\times10^{-5}$~gal$^{-1}$yr$^{-1}$. For rates as a function of MBH mass (right panel), we find remarkable agreement with observations extending well into the IMBH regime.



We are not the first of recent studies to bring theoretical and observed TDE rates into better agreement, and we include these in Figure~\ref{fig:rates_v_mbh_comp} for comparison. \citet{Polkas2024} used semi-analytic galaxy formation and evolution models to measure TDE rate evolution over cosmological timescales while considering galaxy nucleation. We show the $z=0$ results from these simulations as bright green dot-dashed lines in Figure~\ref{fig:rates_v_mbh} (converted from volumetric to per-galaxy rates as in the \citealt{Yao2023} rates). 

Another recent work, \citet{Chang2024}, presents a similar characterization of TDE rates as a function of galaxy mass, but using a very different sample of galaxies.  Rather than focusing on high-quality nuclear stellar density measurements in the nearest galaxies as we do here, this study included two samples of galaxies, (1) 23 galaxies with potential intermediate-mass black holes (IMBHs) both in galaxy nuclei and in massive star clusters, and (2) 37 spiral galaxies with well-measured MBHs from \citet{Davis2019}. Their IMBH candidates include both some with secure dynamical measurements, as well as IMBH candidates with just dynamical upper limits on their masses from \citet{Neumayer2012}.  Thus, the black hole masses in this sample are quite uncertain (as they are in our sample as well).  In many cases, the light profiles for the IMBH hosts are created based solely on a fitted NSC light profile excluding the role of the galaxy.  We note that although the TDE rate is dominated by the NSCs \citep{Pfister2020}, this doesn't mean the depth of the host galaxy potential doesn't impact the TDE rates; we find the inclusion/exclusion of a host galaxy can impact our TDE rates by up to 1 dex.  
The higher mass MBH sample in \citet{Chang2024} has well-determined black hole masses, but the mass profile information comes from low-resolution data relative to our sample and doesn't include nuclear star cluster components \citep{Davis2019}.  We show the individual black holes from \citet{Chang2024} in the right panel of Fig.~\ref{fig:rates_v_mbh_comp} as green crosses; above 10$^6$~M$_\odot$, their galaxies mostly fall below those in our sample. This is likely due to the lack of modeled nuclear star cluster components \citep{Pfister2020}, but may also be due to their focus on massive spiral galaxies, while our sample includes both early and late-type galaxies.  At the low mass end, our data agree fairly well despite the differences noted above.


A novel commonality in our results and those of \citet{Chang2024} is the rollover in per-galaxy TDE rates for IMBHs located in dwarf galactic nuclei.  In contrast to almost all past dynamical modeling of SMBH TDE rates (which decrease with increasing SMBH mass), both \citet{Chang2024} and this paper find that IMBH TDE rates {\it increase} with increasing IMBH mass.  This result is particularly notable insofar as our methodology and samples are quite different as noted above.
We include the average TDE rate as a function of MBH mass from \citet{Chang2024} as a green dashed line in the right panel of Figure~\ref{fig:rates_v_mbh_comp}. Compared to our binned mean fit (blue line) we find a similar shape, but with a significantly offset peak MBH mass, as this was not a fitted parameter in \citet{Chang2024}.
Regardless of these differences, the qualitatively similar conclusions reached in both papers are likely robust to their largest uncertainty (choice of IMBH mass), and the reduced TDE rate in dwarf galactic nuclei likely represents lower density conditions in the NSCs there \citep{Pechetti2020, Hannah2024}. We note that low TDE rates in IMBH nuclei can also be understood from the dashed line in Figure \ref{fig:rates_v_mbh}: if dwarf galactic nuclei had TDE rates $\dot{N}_{\rm TDE} \gtrsim {\rm few} \times 10^{-4}~{\rm yr}^{-1}$, smaller IMBHs would quickly grow into small SMBHs through star capture alone.


Overall, theoretical TDE rates appear to be getting closer to observations. This tentative agreement should be investigated further with new observational TDE rates based on larger samples of TDEs, as opposed to the sample of 33 TDEs used in \citet{Yao2023}. In addition, the previously unobserved class of infrared TDEs found by \citet{Masterson2024} shows a rate similar to optical and X-ray TDEs, suggesting that the existing rate estimates from a single wavelength band may underestimate the total TDE rate. 

\begin{figure}[!t]
    \centering
    \includegraphics[width=0.9\linewidth]{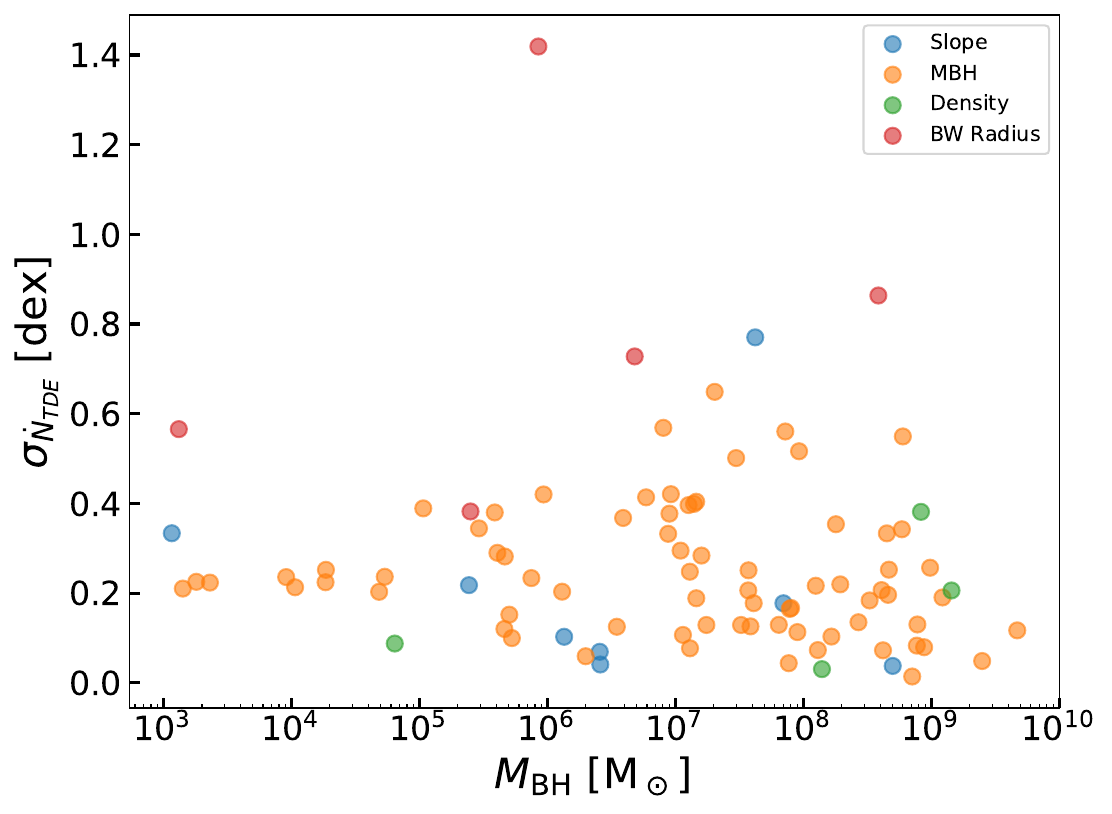}
    \caption{TDE rate uncertainties plotted as a function of MBH mass where the colors indicate the dominate source of error (out of the four explored). While most uncertainties are dominated by errors in MBH mass, the largest uncertainties are due to the sensitivity of steep power-law density profiles to the radial extent of the Bachall-Wolf cusp.}
    \label{fig:uncertainties}
\end{figure}

\begin{figure}[!b]
    \centering
    \includegraphics[width=0.9\linewidth]{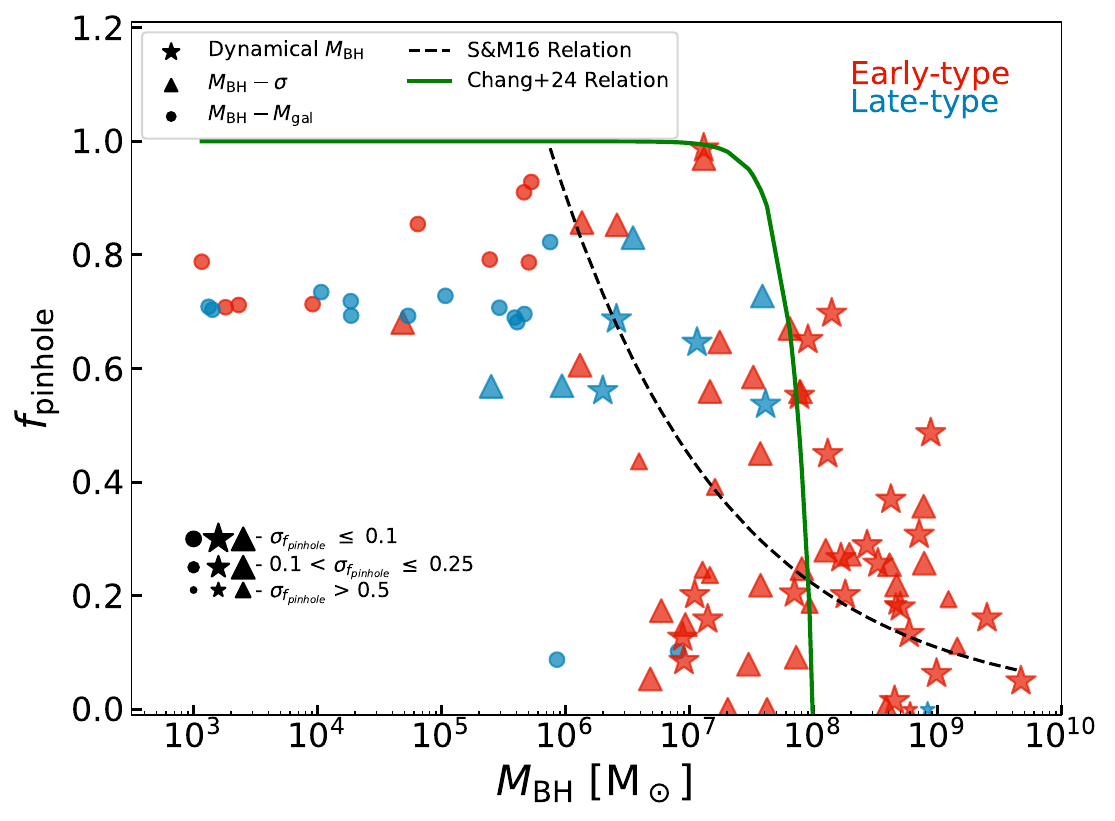}
    \caption{Pinhole fraction ($f_{\rm pinhole}$) for our sample galaxies as a function of MBH mass along with the best-fit relations from \citet{StoneMetzger2016} (black dashed) and \citet{Chang2024} (green).}
    \label{fig:pinhole}
\end{figure}

\subsection{Rate Uncertainties} \label{sec:errors}
To determine the uncertainties on the TDE rates, we explore four primary sources of error: MBH mass, density normalization (largely set by $M/L$ uncertainties), the power-law density slope used in extrapolation, and the radial extent of the Bachall-Wolf cusp, if applied. For the first three sources of error, we perform three rate calculations for each galaxy with the following parameter values: the nominal value, nominal - $1 \sigma$ error, and nominal + $1 \sigma$ error. The errors on MBH masses come from their respective sources, while the uncertainties on the density normalization and slope were derived in \citetalias{Hannah2024}. We then take the standard deviation of each set of three rates to measure the uncertainty originating from each source. 

For steep galaxies with a Bachall-Wolf cusp, we estimate the uncertainty through two rate calculations; one using our nominal definition of the Bachall-Wolf radius based on relaxation over a Hubble time (as discussed in Section~\ref{sec:small_radii}) and another using the observed resolution limit of each galaxy. We define the resulting rate uncertainty as half of the difference in TDE rates between the two cases.

Lastly, these individual errors were added in quadrature to define each galaxy's TDE rate uncertainty (listed in Table~\ref{tab:1}). The individual uncertainties are shown in Figure~\ref{fig:uncertainties} as a function of MBH mass where the symbol colors indicate the dominant source of error. Clearly, most TDE rate uncertainties are driven by the error in the MBH mass measurements. However, some of the largest uncertainties are driven by unknown dynamical ages, which define the radial extent of the Bachall-Wolf cusps applied in galaxies with steep density profiles. The largest TDE rate uncertainty is related to this effect and belongs to NGC~4592. The resolution limit for this galaxy is 2.80~pc while relaxation over a Hubble time infers a Bachall-Wolf cusp radius of 0.09~pc. Although this is not the largest difference in Bachall-Wolf cusp radius between the definitions, it is the steepest galaxy ($\gamma = 2.99$) with such a difference, which leads to the high rate uncertainty.



\subsection{Pinhole Fraction}

The pinhole fraction $f_{\rm pinhole}$ is a common metric presented for TDE rate estimates that describes the fraction of TDEs coming from the ``full loss cone" or "pinhole" regime (i.e. $q > 1$). These TDEs are interesting observationally because the pinhole scenario can produce TDEs with large penetration factors ($\beta = r_{\rm t} / r_{\rm p}$), while in the ``empty loss cone" regime, nearly all TDEs have $\beta \approx 1$ (\citealt{StoneMetzger2016, Broggi2022}, although see also \citealt{Weissbein2017}). This fraction is measured by integrating the loss cone flux (Eq.~\ref{eq:lcflux}) over orbital energies corresponding to $q \leq 1$ and dividing this by the full integral over all energies. 

Figure~\ref{fig:pinhole} shows $f_{\rm pinhole}$ for our sample galaxies as a function of MBH mass. Similarly to previous works, we find that $f_{\rm pinhole}$ increases with decreasing MBH mass. However, we do not observe a nearly 100\% pinhole fraction for $M_\bullet \lesssim 10^6$ as in the parametric fits provided by \citet{Chang2024} and \citet{StoneMetzger2016}, which are shown as the solid green and black dashed lines, respectively. Instead, we observe a convergence toward $f_{\rm pinhole} \approx 0.75$. Our sample contains mostly nucleated galaxies, with all our late-type and low-mass galaxies being nucleated (see \citetalias{Hannah2024} for more details).  This suggests that the high densities of the NSCs in these galaxies may place an upper limit on the pinhole fraction that has not been observed before. 

\subsection{Angular Momentum Diffusion Distributions}

\begin{figure}[!t]
    \centering
    \includegraphics[width=0.9\linewidth]{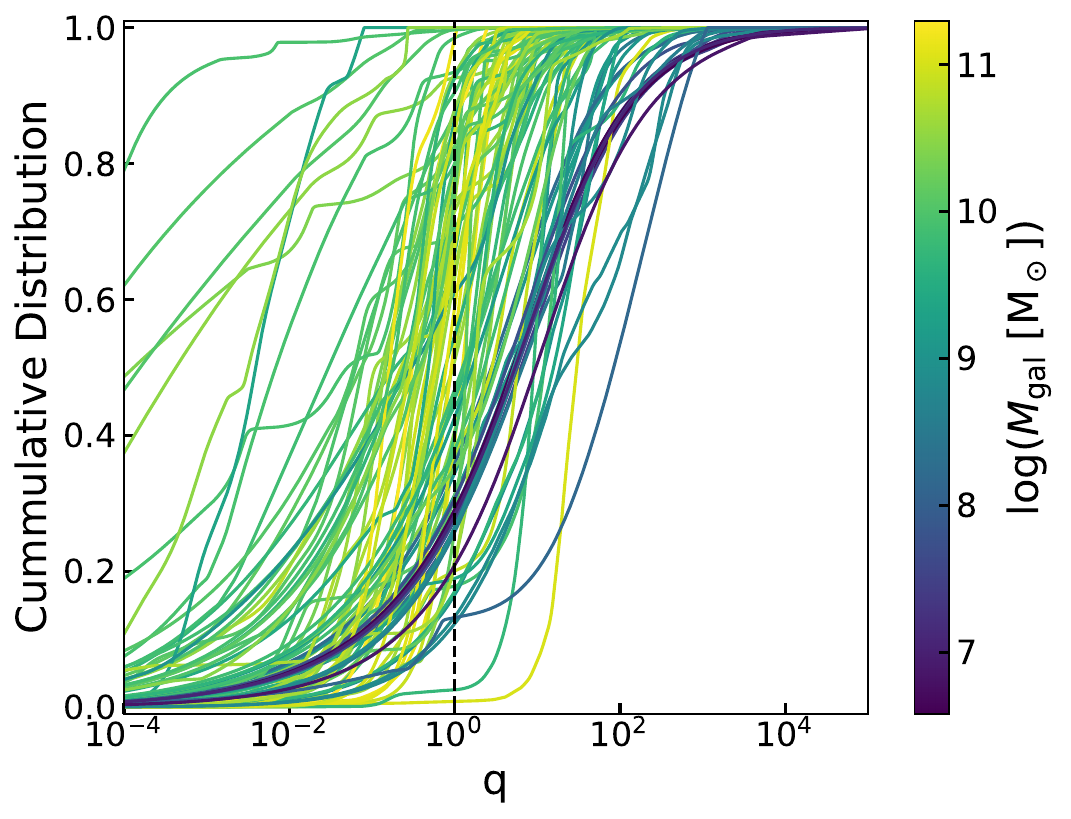}
    \caption{Cumulative $q$ distributions for our sample galaxies highlight the increased contribution to TDEs from the full loss cone regime ($q>1$) in lower mass galaxies. The black dashed line gives the $q = 1$ line.}
    \label{fig:qdists}
\end{figure}

While the pinhole fraction provides some insight into the dynamics of individual TDEs in a galaxy, we take this investigation a step further by deriving the normalized cumulative distributions of $q$ values for each galaxy (see Figure~\ref{fig:qdists}). This is accomplished by multiplying the loss cone flux ($\mathcal{F(\epsilon)} = {\rm d}\dot{N}_{\rm TDE}/{\rm d}\epsilon$) by $|{\rm d}\epsilon/{\rm d}q|$, which we compute with numerical derivatives. These results further highlight the preference for diffusive TDEs (the ``empty loss-cone" regime) in more massive galaxies. Although most TDEs in a galaxy are generally sourced from stars with $q\sim 1$, these distributions highlight subdominant contributions to the rates and how they differ with host galaxy mass. As galaxy mass decreases, the distributions tend to encompass larger $q$ values, visibly highlighting the pinhole TDE preference for lower-mass galaxies. 

\subsection{MBH Spin Evolution}

Assuming an isotropic distribution of impact trajectories, TDEs should serve to spin down an MBH over time \citep{Merritt2013, Metzger2016}, in analogy to the ``chaotic accretion'' paradigm for MBH spin evolution through many AGN episodes \citep{KingPringle06}. Based on this idea, we estimate the maximum dimensionless spin ($a_{\rm max}$) allowed for each of our MBHs.  While in principle, one should calculate $a_{\rm max}$ by combining MBH growth histories (from e.g. mergers and AGN episodes) with a time-dependent TDE rate, here we present a simplified calculation that nonetheless captures basic trends.  We imagine beginning with a maximally spinning ($a_\bullet=.997$) black hole that experiences some mass growth over time due to a constant TDE rate. Under these assumptions, the maximum spin is given by \citep{Metzger2016}:
\begin{equation} \label{eq:amax}
    a_{\rm max} \approx 1- \frac{\dot N_{\rm TDE} \langle M_\star \rangle t_{\rm H}}{2M_\bullet},
\end{equation}
where $t_{\rm H}$ is the age of the universe ($\approx 14 \times 10^9$~yr) and we have assumed that half the star's mass accretes during the TDE.  Note that it is possible in extreme cases for $a_{\rm max} < 0$, in which case we set it to $0$.

\begin{figure}[!t]
    \centering
    \includegraphics[width=0.9\linewidth]{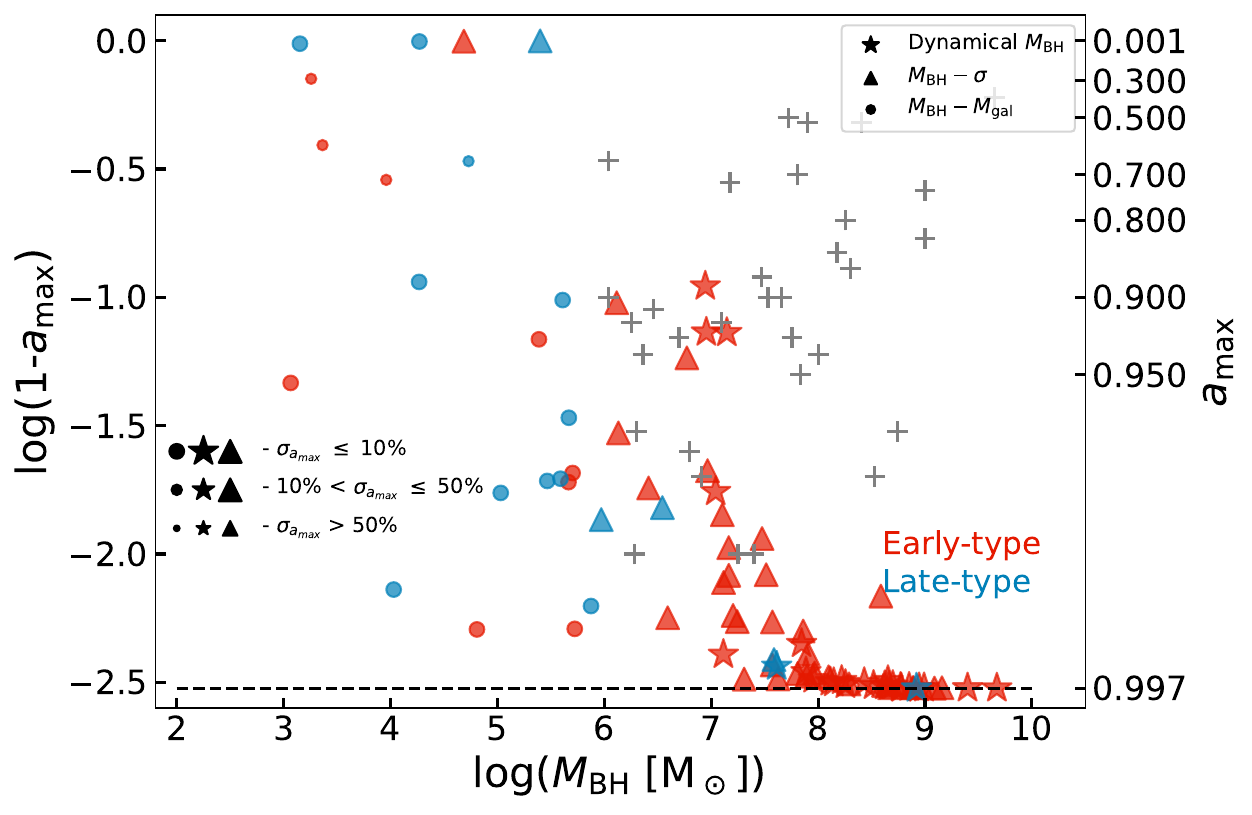}
    \caption{Maximum dimensionless MBH spin values ($a_{\rm max}$) shown as a function of MBH mass, with the theoretical maximum value for thin disk accretion $0.997$ \citep{Thorne74} marked by the black dashed line. The gray crosses represent current SMBH spin measurements from \citet{Reynolds21}.}
    \label{fig:amax}
\end{figure}

Figure~\ref{fig:amax} shows the results of this calculation.  It is clear that spin-down from TDEs is negligible for the highest mass MBHs in our sample. However, TDEs appear to place non-trivial upper limits on the MBH spin for many lower-mass black holes. In particular, for $M_\bullet \lesssim 10^{7} M_\odot$, nuclear TDE rates are generally high enough to prevent time-averaged MBH spin values from reaching the Thorne limit of $a=0.997$ \citep{Thorne74}.  As MBH masses drop, so does (on average) $a_{\rm max}$, sometimes reaching values as low as $\approx 0.9$, well below many astrophysical spin measurements from iron K$\alpha$ spectroscopy \citep[shown as gray crosses;][]{Reynolds21}.  

We note here that $a_{\rm max}$ represents a temporally averaged limit on MBH spin.  If indeed one finds (via X-ray observations or other techniques) a relatively small MBH ($\lesssim 10^5 M_\odot$) with very large $a$, that could indicate (i) an abnormally low TDE rate in the galaxy, or, more likely (ii) that the galaxy has grown dramatically from an ongoing episode of {\it coherent} accretion, more than doubling its mass to spin up beyond the temporally averaged upper limit provided by TDEs.  The existence of a preferred direction of TDE angular momentum (e.g. a nuclear star cluster with net rotation) could also weaken this constraint, but except in the extreme case of a disk-like cluster, should only affect $1-a_{\rm max}$ by a factor of order unity.

\section{Conclusions \& Future Work} 
\label{sec:con-future}

In this work, we have developed a new software package, REPTiDE, for computing 2-body relaxation-based TDE rates given a galaxy density profile and MBH mass. The output from this code agrees well with previous implementations of the loss cone procedure and will be a useful tool for future TDE rate studies enabled by the next generation of surveys. We demonstrate here that REPTiDE agrees well with the  \citet{StoneMetzger2016} implementation of the standard loss cone theory (Figure~\ref{fig:accuracy}). While developed primarily for application to large samples of model galaxies, in this work we applied REPTiDE to the real, observed galaxies used to construct the new density scaling relations presented in \citetalias{Hannah2024}. 

Given our strict nuclear density requirements during sample construction (minimum radial resolution of 5~pc), the rate calculations presented in this paper represent the most accurate individual TDE rate estimations for a sample of nucleated galaxies. We found that our TDE rate distributions peak around log($M_{\rm gal}/{\rm M}_\odot$)$ \approx 9.4$ and log($M_\bullet/{\rm M}_\odot$)$ \approx 6.5$. We fit broken power laws directly to these distributions resulting in relations that describe the typical TDE rates as functions of galaxy and MBH mass.

The large variance in TDE rates between galaxies of similar properties (e.g. mass, or MBH mass) means that observed TDE rates will often be dominated by the galaxy subpopulations with the highest TDE rates. Additionally, TDE rate suppression due to direct capture heavily alters the observability of TDEs at the high mass end. We, therefore, fitted a double broken power law to the binned mean of each distribution with rates that include event horizon suppression. We compared these with the most recent observed TDE rates from \citet{Yao2023} and found excellent agreement. At galaxy masses $\lesssim 10^{10.5}$~M$_\odot$, this sample is composed entirely of NSC-hosting galaxies (for both early- and late-types). Thus, achieving this agreement with a limited sample of nearby nucleated galaxies further highlights the importance of considering NSC density contributions when modeling TDE rates \citep[e.g.][]{Pfister2020}. 
 
We also presented the pinhole fraction for each of our sample galaxies, which describes the fraction of TDEs resulting from the ``full" loss cone (pinhole) regime ($q>1$). Pinhole regime TDEs generally result in full disruptions while diffusive regime events (i.e. $q<1$) can produce overwhelmingly partial TDEs (where the star is not fully destroyed on a single passage; \citealt{Broggi2024}). The distributions of individual TDE parameters in a galaxy could impact the observed TDE rate. In agreement with previous works \citep[i.e.][]{StoneMetzger2016, Chang2024}, we find that the pinhole fraction increases rapidly with decreasing MBH mass (Figure~\ref{fig:pinhole}), possibly biasing larger SMBHs against producing (more easily observable) full disruptions. However, unlike in past works, we observed a convergence in the pinhole fraction to $\approx0.75$ at low MBH masses, which we attributed to the elevated densities in NSCs causing a non-negligble fraction of TDEs to be sourced from very near the MBH. This has not been observed before, and should be tested against a sample of non-nucleated low-mass galaxies with similar density resolution. 

The normalized cumulative $q$ distributions for each galaxy (Figure~\ref{fig:qdists}) shows a more detailed view of the preference for pinhole TDEs in low-mass galaxies. They also highlight how the overall ranges of $q$ values contributing to the TDE rates change with galaxy mass, where higher-mass galaxies produce TDEs down to much lower $q$ values. 

Lastly, we investigated the ability for TDEs to place upper limits on MBH spin ($a_\bullet$), assuming that the quasi-isotropic nature of stellar orbital inclinations will serve to spin down an MBH. Assuming a near-maximal MBH spin \citep[$a_\bullet=0.997$;][]{Thorne74}, we compute the maximum allowed spin ($a_{\rm max}$) for each MBH (Equation~\ref{eq:amax}) assuming a constant TDE rate over a Hubble time (Figure~\ref{fig:amax}). While TDE rates are expected to evolve with time, comparing these static calculations with other MBH spin measurements could place constraints on the time evolution of the TDE rate in a galaxy.  AGN spin measurements that find $a_\bullet$ values far above $a_{\rm max}$ can indicate ongoing episodes of coherent accretion.

In general, we do not find evidence for the ``rate discrepancy'' identified by some past semi-empirical loss cone modeling \citep{Wang2004, StoneMetzger2016}, i.e. a mismatch between (higher) theoretical rates and (lower) observed rates.  Although it is possible that physical changes to simpler loss cone models could reduce dynamical rates \citep{Lezhnin2015, Teboul2024, Broggi2024, Teboul2024b}, our underlying dynamical treatment is broadly the same as in past semi-empirical work \citep{StoneMetzger2016}: the only thing that has changed is the underlying sample of galaxies we average over.  We believe that this highlights the importance of constructing theoretical (per-galaxy or volumetric) TDE rate estimates from the most {\it representative} galaxy samples possible, something that is challenging to do using archival {\it HST} data alone because of the {\it ad hoc} selection criteria that have generally determined the availability of high-resolution nuclear {\it HST} photometry.  One solution to this problem is to rely on a purely theoretical catalog of galaxies and nuclear dynamics \citep{Polkas2024}, but it is also possible to retain the direct link to observed nuclear dynamics.

In Paper III (Hannah et al. {\it in-prep}), REPTiDE will be applied to model galaxy samples constructed with existing galaxy scaling relations as well as the new NSC density scaling relations from \citetalias{Hannah2024}. Each sample will be created with varying MBH occupation fractions and mass distributions based on expectations from different MBH formation scenarios. The intrinsic TDE rates computed for each model sample will be forward modeled into detection rates for ZTF, Rubin, and {\it ULTRASAT}. As ZTF is the current leader in TDE detections, immediate comparisons with the observed rates will be possible. However, ZTF samples are still small ($\lesssim100$), so constraints may be limited in this case. Thus, we will forward model our predictions for Rubin and {\it ULTRASAT} as well, enabling future rate comparisons for expected near-future samples of thousands of observed TDEs.

\section*{Acknowledgements}
C.H.H. and A.C.S. acknowledge support from National Science Foundation astronomy and astrophysics grant AST-2108180. N.C.S. acknowledges financial support from the Israel Science Foundation (Individual Research Grant 2565/19) and the Binational Science Foundation (grant Nos. 2019772 and 2020397).

\bibliography{references}
\bibliographystyle{aasjournal}

\startlongtable
\begin{deluxetable}{ccccccccccc}
\tablecolumns{11}
\tabletypesize{\scriptsize}
\tablecaption{Tabulated TDE rate results for all 91 galaxies. \label{tab:1}}
\tablehead{
    \colhead{Name} & \colhead{Type} & \colhead{log(M$_{\rm{gal}}$)} & \colhead{log(M$_{\rm{BH}}$)} & \colhead{source} & \colhead{$\gamma$} & \colhead{log($\rho_{5pc}$)} & \colhead{$\dot N_{\rm TDE}^{\rm EHS}$} & \colhead{$\dot N_{\rm TDE}$} & \colhead{BW Cusp} & \colhead{$r_{\rm BW}$} \\
    \colhead{(1)} & \colhead{(2)} & \colhead{(3)} & \colhead{(4)} & \colhead{(5)} & \colhead{(6)} & \colhead{(7)} & \colhead{(8)} & \colhead{(9)} &
    \colhead{(10)} & \colhead{(11)}
}

\startdata
\hline
BTS 076 & Late & 7.51 & $4.03^{+0.60}_{-0.60}$ & 5 & -2.41 & 0.41 & $-7.66 \pm 0.21$ & $-7.66 \pm 0.21$ & True & 2.14e+00 \\ 
BTS 109 & Late & 6.58 & $3.12^{+0.60}_{-0.60}$ & 5 & -2.97 & 0.15 & $-6.4 \pm 0.57$ & $-6.4 \pm 0.57$ & True & 6.92e-01 \\ 
DDO 084 & Late & 8.53 & $5.03^{+0.60}_{-0.60}$ & 5 & -2.9 & 1.1 & $-6.14 \pm 0.39$ & $-6.14 \pm 0.39$ & True & 9.50e-01 \\ 
ESO 274-1 & Late & 9.12 & $5.61^{+0.60}_{-0.60}$ & 5 & -2.71 & 2.3 & $-4.74 \pm 0.29$ & $-4.74 \pm 0.29$ & True & 4.73e-01 \\ 
IC 5052 & Late & 9.18 & $5.67^{+0.60}_{-0.60}$ & 5 & -2.42 & 2.4 & $-5.16 \pm 0.28$ & $-5.16 \pm 0.28$ & True & 9.33e-01 \\ 
LeG 09 & Early & 6.99 & $3.26^{+0.65}_{-0.65}$ & 5 & -2.15 & 1.24 & $-6.22 \pm 0.23$ & $-6.22 \pm 0.23$ & True & 1.73e+00 \\ 
NGC 0584 & Early & 10.56 & $8.11^{+0.14}_{-0.21}$ & 3 & -1.38 & 3.74 & - & $-4.97 \pm 0.07$ & False & - \\ 
NGC 0596 & Early & 10.46 & $7.89^{+0.43}_{-0.43}$ & 4 & -1.45 & 3.75 & $-5.83 \pm 0.16$ & $-4.89 \pm 0.16$ & False & - \\ 
NGC 0821 & Early & 10.7 & $8.22^{+0.21}_{-0.21}$ & 1 & -1.36 & 4.03 & - & $-4.78 \pm 0.10$ & False & - \\ 
NGC 1374 & Early & 10.3 & $8.77^{+0.32}_{-1.07}$ & 3 & -0.4 & 3.64 & - & $-5.27 \pm 0.34$ & False & - \\ 
NGC 1399 & Early & 11.16 & $8.94^{+0.48}_{-0.30}$ & 3 & -1.12 & 2.63 & - & $-5.46 \pm 0.08$ & False & - \\ 
NGC 1427 & Early & 10.39 & $7.97^{+0.43}_{-0.43}$ & 4 & -2.24 & 3.7 & $-6.0 \pm 0.52$ & $-4.71 \pm 0.52$ & True & 6.82e-03 \\ 
NGC 1439 & Early & 10.49 & $7.86^{+0.43}_{-0.43}$ & 4 & -2.23 & 4.02 & $-4.99 \pm 0.56$ & $-4.16 \pm 0.56$ & True & 2.89e-02 \\ 
NGC 2300 & Early & 10.81 & $9.09^{+0.43}_{-0.43}$ & 4 & -0.82 & 2.47 & - & $-5.68 \pm 0.19$ & False & - \\ 
NGC 2434 & Early & 10.58 & $8.29^{+0.43}_{-0.43}$ & 4 & -1.15 & 3.98 & - & $-4.96 \pm 0.22$ & False & - \\ 
NGC 2778 & Early & 9.94 & $7.15^{+0.49}_{-0.54}$ & 3 & -2.05 & 4.29 & $-3.33 \pm 0.40$ & $-3.33 \pm 0.40$ & True & 5.30e-01 \\ 
NGC 2787 & Late & 9.95 & $7.61^{+0.32}_{-0.91}$ & 3 & -1.31 & 3.7 & $-5.24 \pm 0.18$ & $-4.9 \pm 0.18$ & False & - \\ 
NGC 2903 & Late & 10.4 & $7.06^{+0.28}_{-7.06}$ & 1 & -1.65 & 3.45 & $-5.0 \pm 0.11$ & $-5.0 \pm 0.11$ & False & - \\ 
NGC 3115B & Early & 9.03 & $4.69^{+0.43}_{-0.43}$ & 4 & -1.85 & 2.9 & $-4.5 \pm 0.20$ & $-4.5 \pm 0.20$ & True & 1.65e+00 \\ 
NGC 3274 & Late & 8.22 & $4.73^{+0.60}_{-0.60}$ & 5 & -2.84 & 1.92 & $-5.07 \pm 0.24$ & $-5.07 \pm 0.24$ & True & 1.18e+00 \\ 
NGC 3344 & Late & 9.9 & $6.54^{+0.50}_{-0.50}$ & 4 & -1.62 & 3.49 & $-4.69 \pm 0.12$ & $-4.69 \pm 0.12$ & False & - \\ 
NGC 3379 & Early & 10.63 & $8.62^{+0.35}_{-0.58}$ & 3 & -1.08 & 3.51 & - & $-4.9 \pm 0.07$ & False & - \\ 
NGC 3384 & Early & 10.13 & $7.04^{+0.39}_{-0.33}$ & 3 & -1.98 & 3.75 & $-4.12 \pm 0.29$ & $-4.12 \pm 0.29$ & True & 1.12e-01 \\ 
NGC 3412 & Early & 10.01 & $7.31^{+0.43}_{-0.43}$ & 4 & -2.23 & 2.87 & $-5.66 \pm 0.65$ & $-5.61 \pm 0.65$ & True & 9.28e-03 \\ 
NGC 3522 & Early & 9.69 & $7.1^{+0.43}_{-0.43}$ & 4 & -1.94 & 3.73 & $-4.17 \pm 0.40$ & $-4.17 \pm 0.40$ & True & 4.41e-02 \\ 
NGC 3585 & Early & 10.93 & $8.52^{+0.38}_{-0.74}$ & 3 & -1.23 & 3.8 & - & $-4.94 \pm 0.18$ & False & - \\ 
NGC 3607 & Early & 10.87 & $8.15^{+0.36}_{-0.45}$ & 3 & -0.95 & 3.48 & - & $-5.0 \pm 0.03$ & False & - \\ 
NGC 3608 & Early & 10.45 & $8.66^{+0.35}_{-0.71}$ & 3 & -0.98 & 3.74 & - & $-4.88 \pm 0.20$ & False & - \\ 
NGC 3610 & Early & 10.8 & $8.1^{+0.43}_{-0.43}$ & 4 & -1.77 & 4.05 & - & $-4.72 \pm 0.22$ & True & 1.27e-07 \\ 
NGC 3640 & Early & 10.9 & $7.89^{+0.22}_{-0.46}$ & 3 & -0.23 & 3.56 & $-5.65 \pm 0.04$ & $-4.76 \pm 0.04$ & False & - \\ 
NGC 3945 & Early & 10.54 & $6.94^{+0.60}_{-0.05}$ & 3 & -2.4 & 4.11 & $-3.35 \pm 0.33$ & $-3.35 \pm 0.33$ & True & 7.80e-01 \\ 
NGC 4026 & Early & 10.06 & $8.26^{+0.37}_{-0.65}$ & 3 & -1.67 & 3.93 & - & $-5.07 \pm 0.35$ & False & - \\ 
NGC 4242 & Late & 9.1 & $5.59^{+0.60}_{-0.60}$ & 5 & -3.07 & 1.67 & $-5.51 \pm 0.38$ & $-5.51 \pm 0.38$ & True & 1.20e+00 \\ 
NGC 4262 & Early & 9.97 & $8.61^{+0.43}_{-0.43}$ & 4 & -1.32 & 4.17 & - & $-4.81 \pm 0.21$ & False & - \\ 
NGC 4291 & Early & 10.5 & $8.99^{+0.37}_{-0.50}$ & 3 & -0.56 & 2.91 & - & $-5.45 \pm 0.26$ & False & - \\ 
NGC 4342 & Early & 9.93 & $8.65^{+0.41}_{-0.48}$ & 3 & -1.99 & 4.48 & - & $-4.35 \pm 0.33$ & True & 6.26e-05 \\ 
NGC 4365 & Early & 11.09 & $8.89^{+0.43}_{-0.43}$ & 4 & -0.5 & 2.97 & - & $-5.28 \pm 0.08$ & False & - \\ 
NGC 4377 & Early & 10.02 & $7.62^{+0.43}_{-0.43}$ & 4 & -2.29 & 3.16 & $-5.64 \pm 0.77$ & $-5.28 \pm 0.77$ & True & 9.28e-03 \\ 
NGC 4379 & Early & 9.92 & $7.48^{+0.43}_{-0.43}$ & 4 & -2.32 & 3.96 & $-4.1 \pm 0.50$ & $-3.92 \pm 0.50$ & True & 1.85e-01 \\ 
NGC 4382 & Early & 11.02 & $7.11^{+1.25}_{-0.64}$ & 3 & -0.57 & 2.99 & $-5.18 \pm 0.25$ & $-5.18 \pm 0.25$ & False & - \\ 
NGC 4387 & Early & 9.84 & $7.16^{+0.43}_{-0.43}$ & 4 & -2.15 & 3.75 & $-4.28 \pm 0.40$ & $-4.28 \pm 0.40$ & True & 1.68e-01 \\ 
NGC 4434 & Early & 10.03 & $7.85^{+0.13}_{-0.19}$ & 3 & -1.87 & 4.17 & $-5.11 \pm 0.18$ & $-4.31 \pm 0.18$ & True & 3.07e-04 \\ 
NGC 4458 & Early & 9.73 & $7.12^{+0.43}_{-0.43}$ & 4 & -0.62 & 3.79 & $-4.53 \pm 0.08$ & $-4.53 \pm 0.08$ & False & - \\ 
NGC 4464 & Early & 9.58 & $7.57^{+0.43}_{-0.43}$ & 4 & -1.53 & 3.7 & $-5.19 \pm 0.21$ & $-4.91 \pm 0.21$ & False & - \\ 
NGC 4467 & Early & 9.04 & $6.41^{+0.43}_{-0.43}$ & 4 & -1.33 & 3.58 & $-4.73 \pm 0.04$ & $-4.73 \pm 0.04$ & False & - \\ 
NGC 4472 & Early & 11.3 & $9.4^{+0.35}_{-1.40}$ & 3 & -1.46 & 2.76 & - & $-5.31 \pm 0.05$ & False & - \\ 
NGC 4473 & Early & 10.49 & $7.95^{+0.41}_{-0.30}$ & 3 & -0.81 & 3.52 & $-6.16 \pm 0.11$ & $-4.96 \pm 0.11$ & False & - \\ 
NGC 4474 & Early & 9.97 & $6.97^{+0.43}_{-0.43}$ & 4 & -2.18 & 3.61 & $-4.1 \pm 0.42$ & $-4.1 \pm 0.42$ & True & 3.08e-01 \\ 
NGC 4478 & Early & 10.08 & $7.81^{+0.43}_{-0.43}$ & 4 & -1.79 & 3.61 & $-5.57 \pm 0.13$ & $-4.89 \pm 0.13$ & True & 2.67e-07 \\ 
NGC 4483 & Early & 9.72 & $7.16^{+0.43}_{-0.43}$ & 4 & -1.66 & 3.87 & $-4.44 \pm 0.19$ & $-4.44 \pm 0.19$ & False & - \\ 
NGC 4486B & Early & 9.33 & $8.78^{+0.40}_{-0.48}$ & 3 & -1.0 & 3.46 & - & $-6.59 \pm 0.55$ & False & - \\ 
NGC 4489 & Early & 9.65 & $6.13^{+0.43}_{-0.43}$ & 4 & -1.35 & 3.42 & $-4.77 \pm 0.10$ & $-4.77 \pm 0.10$ & False & - \\ 
NGC 4494 & Early & 10.68 & $7.91^{+0.43}_{-0.43}$ & 4 & -1.44 & 4.24 & $-5.36 \pm 0.17$ & $-4.38 \pm 0.17$ & False & - \\ 
NGC 4517 & Late & 9.86 & $5.97^{+0.50}_{-0.50}$ & 4 & -2.72 & 2.12 & $-5.33 \pm 0.42$ & $-5.33 \pm 0.42$ & True & 7.85e-01 \\ 
NGC 4528 & Early & 9.85 & $7.24^{+0.43}_{-0.43}$ & 4 & -1.55 & 3.71 & $-4.69 \pm 0.13$ & $-4.69 \pm 0.13$ & False & - \\ 
NGC 4551 & Early & 9.86 & $7.2^{+0.43}_{-0.43}$ & 4 & -1.88 & 3.62 & $-4.68 \pm 0.28$ & $-4.68 \pm 0.28$ & True & 4.02e-03 \\ 
NGC 4552 & Early & 10.7 & $8.7^{+0.05}_{-0.05}$ & 3 & -2.15 & 3.52 & - & $-4.77 \pm 0.04$ & True & 3.42e-05 \\ 
NGC 4589 & Early & 10.72 & $8.67^{+0.43}_{-0.43}$ & 4 & -0.79 & 3.33 & - & $-5.61 \pm 0.25$ & False & - \\ 
NGC 4592 & Late & 9.45 & $5.93^{+0.60}_{-0.60}$ & 5 & -2.99 & 2.33 & $-2.89 \pm 1.42$ & $-2.89 \pm 1.42$ & True & 8.74e-02 \\ 
NGC 4600 & Early & 9.01 & $6.59^{+0.43}_{-0.43}$ & 4 & -2.02 & 2.83 & $-5.31 \pm 0.37$ & $-5.31 \pm 0.37$ & True & 4.65e-02 \\ 
NGC 4605 & Late & 9.17 & $5.4^{+0.50}_{-0.50}$ & 4 & -2.9 & 2.97 & $-3.0 \pm 0.38$ & $-3.0 \pm 0.38$ & True & 4.18e-01 \\ 
NGC 4612 & Early & 9.97 & $6.77^{+0.43}_{-0.43}$ & 4 & -2.53 & 3.7 & $-3.81 \pm 0.41$ & $-3.81 \pm 0.41$ & True & 7.64e-01 \\ 
NGC 4623 & Early & 9.71 & $6.68^{+0.43}_{-0.43}$ & 4 & -2.85 & 3.58 & $-3.08 \pm 0.73$ & $-3.08 \pm 0.73$ & True & 3.87e-01 \\ 
NGC 4638 & Early & 10.14 & $7.51^{+0.43}_{-0.43}$ & 4 & -1.36 & 4.37 & $-4.3 \pm 0.13$ & $-4.08 \pm 0.13$ & False & - \\ 
NGC 4649 & Early & 11.18 & $9.67^{+0.35}_{-0.67}$ & 3 & -1.47 & 3.02 & - & $-5.43 \pm 0.12$ & False & - \\ 
NGC 4660 & Early & 9.93 & $8.59^{+0.43}_{-0.43}$ & 4 & -2.65 & 4.57 & - & $-3.15 \pm 0.86$ & True & 7.15e-02 \\ 
NGC 4733 & Early & 9.83 & $6.12^{+0.43}_{-0.43}$ & 4 & -3.98 & 3.24 & $-4.24 \pm 0.20$ & $-4.24 \pm 0.20$ & True & 2.95e+00 \\ 
NGC 5011C & Early & 7.52 & $3.96^{+0.65}_{-0.65}$ & 5 & -2.63 & 0.98 & $-5.91 \pm 0.24$ & $-5.91 \pm 0.24$ & True & 6.33e-01 \\ 
NGC 5055 & Late & 10.5 & $8.92^{+0.10}_{-0.10}$ & 1 & -1.66 & 3.58 & - & $-6.67 \pm 0.38$ & False & - \\ 
NGC 5068 & Late & 9.39 & $5.88^{+0.60}_{-0.60}$ & 5 & -1.84 & 2.27 & $-5.93 \pm 0.23$ & $-5.93 \pm 0.23$ & True & 7.44e-02 \\ 
NGC 5195 & Early & 10.1 & $7.57^{+0.43}_{-0.43}$ & 4 & -1.77 & 4.09 & $-4.65 \pm 0.25$ & $-4.36 \pm 0.25$ & True & 1.03e-04 \\ 
NGC 5236 & Late & 10.44 & $6.91^{+0.60}_{-0.60}$ & 5 & -2.54 & 3.95 & $-3.28 \pm 0.57$ & $-3.28 \pm 0.57$ & True & 7.06e-01 \\ 
NGC 5238 & Late & 7.75 & $4.27^{+0.60}_{-0.60}$ & 5 & -2.94 & 1.59 & $-4.69 \pm 0.25$ & $-4.69 \pm 0.25$ & True & 5.90e-01 \\ 
NGC 5457 & Late & 10.21 & $6.41^{+0.08}_{-6.41}$ & 1 & -1.81 & 3.05 & $-5.06 \pm 0.07$ & $-5.06 \pm 0.07$ & True & 5.33e-02 \\ 
NGC 5557 & Early & 10.88 & $9.16^{+0.43}_{-0.43}$ & 4 & -0.72 & 3.39 & - & $-5.05 \pm 0.21$ & False & - \\ 
NGC 5576 & Early & 10.58 & $8.43^{+0.35}_{-0.53}$ & 3 & -1.11 & 4.0 & - & $-4.71 \pm 0.14$ & False & - \\ 
NGC 5813 & Early & 11.08 & $8.85^{+0.05}_{-0.06}$ & 3 & -0.01 & 2.37 & - & $-5.13 \pm 0.01$ & False & - \\ 
NGC 5982 & Early & 10.97 & $8.89^{+0.43}_{-0.43}$ & 4 & -0.53 & 2.68 & - & $-5.31 \pm 0.13$ & False & - \\ 
NGC 6503 & Late & 9.64 & $6.3^{+0.11}_{-6.30}$ & 1 & -2.32 & 3.06 & $-4.36 \pm 0.06$ & $-4.36 \pm 0.06$ & True & 9.94e-01 \\ 
NGC 7457 & Early & 9.88 & $6.95^{+0.41}_{-0.22}$ & 3 & -2.58 & 3.79 & $-3.52 \pm 0.38$ & $-3.52 \pm 0.38$ & True & 6.27e-01 \\ 
NGC 7713 & Late & 8.97 & $5.47^{+0.60}_{-0.60}$ & 5 & -3.36 & 1.5 & $-5.65 \pm 0.34$ & $-5.65 \pm 0.34$ & True & 1.32e+00 \\ 
NGC 7727 & Late & 10.67 & $7.59^{+0.50}_{-0.50}$ & 4 & -1.29 & 3.68 & $-5.1 \pm 0.13$ & $-4.79 \pm 0.13$ & False & - \\ 
PGC 4310323 & Late & 6.61 & $3.15^{+0.60}_{-0.60}$ & 5 & -2.76 & 0.87 & $-5.98 \pm 0.21$ & $-5.98 \pm 0.21$ & True & 1.09e+00 \\ 
UGC 07242 & Late & 7.75 & $4.27^{+0.60}_{-0.60}$ & 5 & -2.88 & 1.05 & $-6.01 \pm 0.22$ & $-6.01 \pm 0.22$ & True & 9.25e-01 \\ 
VCC 1199 & Early & 8.6 & $5.39^{+0.65}_{-0.65}$ & 5 & -1.59 & 2.99 & $-5.12 \pm 0.22$ & $-5.12 \pm 0.22$ & False & - \\ 
VCC 1440 & Early & 8.83 & $5.7^{+0.65}_{-0.65}$ & 5 & -1.61 & 2.85 & $-5.37 \pm 0.15$ & $-5.37 \pm 0.15$ & False & - \\ 
VCC 1545 & Early & 8.85 & $5.72^{+0.65}_{-0.65}$ & 5 & -1.23 & 2.37 & $-6.27 \pm 0.10$ & $-6.27 \pm 0.10$ & False & - \\ 
VCC 1627 & Early & 8.8 & $5.66^{+0.65}_{-0.65}$ & 5 & -1.27 & 3.08 & $-5.45 \pm 0.12$ & $-5.45 \pm 0.12$ & False & - \\\relax 
[KK2000] 03 & Early & 8.16 & $4.81^{+0.65}_{-0.65}$ & 5 & -1.84 & 1.21 & $-7.2 \pm 0.09$ & $-7.2 \pm 0.09$ & True & 3.39e-01 \\\relax 
[KK2000] 53 & Early & 6.85 & $3.07^{+0.65}_{-0.65}$ & 5 & -1.65 & 0.84 & $-7.62 \pm 0.33$ & $-7.62 \pm 0.33$ & False & - \\\relax 
[KK98] 096 & Early & 7.07 & $3.36^{+0.65}_{-0.65}$ & 5 & -2.31 & 1.1 & $-6.37 \pm 0.22$ & $-6.37 \pm 0.22$ & True & 1.70e+00 \\ 
\enddata

\tablecomments{(1) galaxy name, (2) galaxy type, (3) galaxy stellar mass [M$_\odot$], (4) black hole mass [M$_\odot$], (5) source of MBH mass: 1=\citet{van-den-Bosch2016}, 2=\citet{Reines2015}, 3=\citet{Greene2020}, 4=\citet{Greene2020} M$_{BH}$-$\sigma$ relations, 5=\citet{Greene2020} M$_{BH}$-M$_{gal}$ relations,
(6) \& (7) give the power-law fit parameters ($\gamma$=slope, $\rho_{5pc}$=3D stellar density at r=5pc [M$_\odot$/pc$^3$]) to the nuclear density profiles from \citetalias{Hannah2024}, (8) full TDE rate with event horizon suppression, (9) full TDE rate without event horizon suppression, (11) flag to indicate if a Bachall-Wolf cusp was applied to the density profile at small radii, (12) radius at which Bachall-Wolf cusp begins [pc].}

\end{deluxetable}

\restartappendixnumbering
\appendix
\section{Orbital Energy Convergence Test}
\label{app:converge}

\begin{figure}
    \centering
    \includegraphics[width=0.9\linewidth]{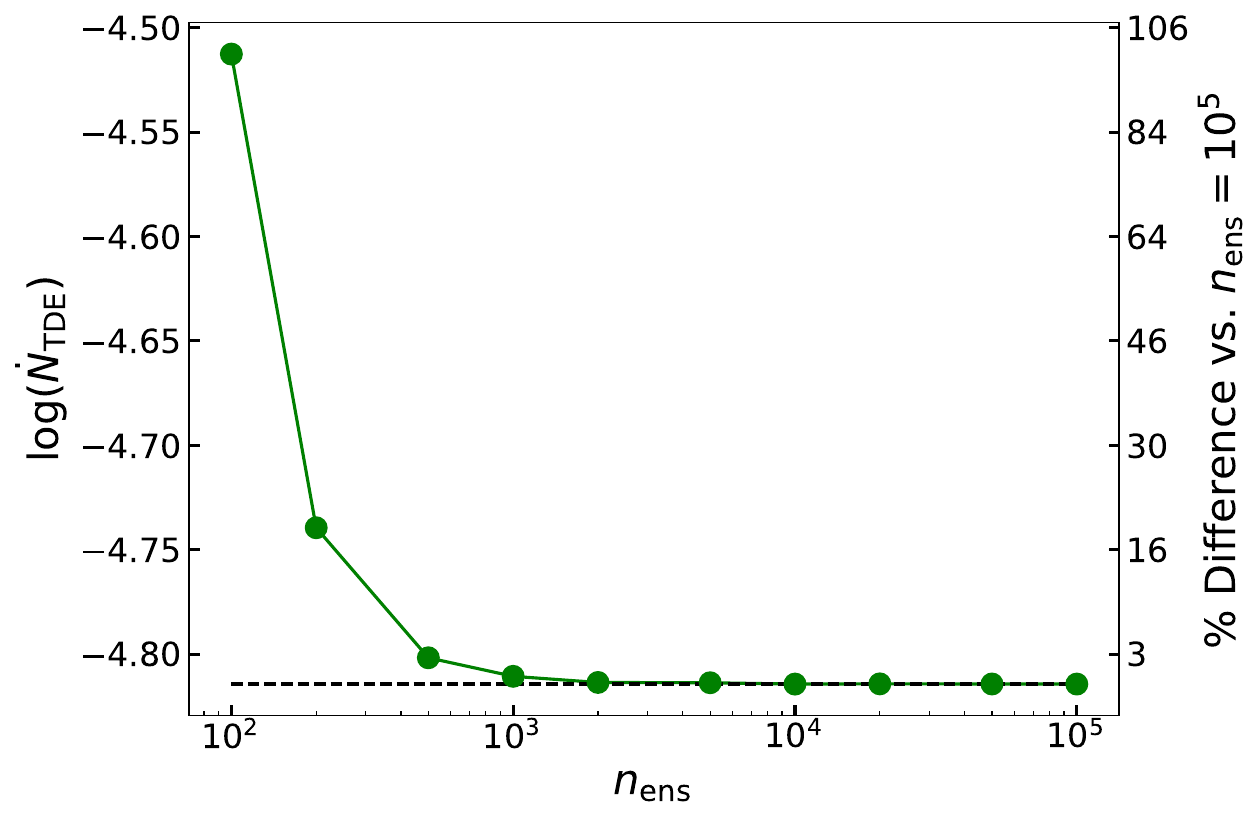}
    \caption{TDE rate as a function of orbital energy grid density specified by the number of energies $n_{\rm ens}$. The right y-axis is labeled to show the percent difference in TDE rates when compared to the densest grid ($n_{\rm ens}$ = $10^5$).} 
    \label{fig:convergence}
\end{figure}

The number of orbital energies ($n_{\rm ens}$) considered when calculating the TDE rate is essentially a free parameter that is directly linked to the runtime. Therefore, we performed a convergence test for TDE rates as a function of ($n_{\rm ens}$) to minimize execution time while maintaining accuracy. Figure~\ref{fig:convergence} shows the results of this investigation for NGC 4551. 

\end{document}